\documentclass[a4paper, 11pt]{article}
\usepackage[utf8]{inputenc}
\usepackage[english]{babel}
\usepackage[T1]{fontenc}

\usepackage[a4paper, top=3cm, bottom=3cm, left=2.5cm, right=2.5cm, marginparwidth=2cm]{geometry}

\usepackage{amsmath}
\usepackage{amsfonts}
\usepackage{graphicx}
\usepackage{caption}
\usepackage{subcaption}
\usepackage[colorlinks=true, allcolors=blue]{hyperref}
\usepackage{setspace}
\usepackage{float}

\usepackage{makecell}
\usepackage{booktabs}
\usepackage[table, dvipsnames]{xcolor} 
\usepackage{multirow} 

\usepackage{doi, natbib, har2nat}
\providecommand{\keywords}[1]
{
	\small	
	\textbf{\textit{Keywords:}} #1
}
\providecommand{\classcodes}[1]
{
	\small	
	\textbf{\textit{JEL classification:}} #1
}

\newcommand{\quantlet}{\raisebox{-1pt}{\protect \includegraphics[scale=0.05]{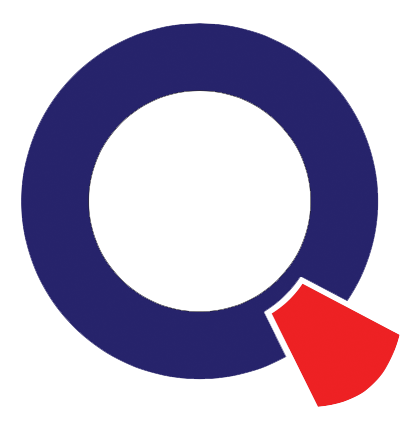}}\,}

\definecolor{seed-node}{HTML}{E7E276}
\definecolor{author-node}{HTML}{EC594D}
\definecolor{course-node}{HTML}{5E974D}
\definecolor{courselet-node}{HTML}{97984F}
\definecolor{order-node}{HTML}{3745F6}
\definecolor{review-node}{HTML}{FBD1D2}

\title{Quantinar: a blockchain peer-to-peer ecosystem for modern data analytics}
\author{
	Raul Bag \footnote{corresponding author, IRTG 1792, School of Business and Economics, Humboldt-Universit{\"a}t zu Berlin, Dorotheenstr. 1, 10117 Berlin, Germany. Email: bagcrist@hu-berlin.de},
	Bruno Spilak \footnote{IRTG 1792, School of Business and Economics, Humboldt-Universit{\"a}t zu Berlin, Dorotheenstr. 1, 10117 Berlin, Germany},
	Julian Winkel \footnote{IRTG 1792, School of Business and Economics, Humboldt-Universit{\"a}t zu Berlin, Dorotheenstr. 1, 10117 Berlin, Germany},
	Wolfgang Karl H{\"a}rdle \footnote{Humboldt-Universität zu Berlin, BRC Blockchain Research Center, Berlin; IRTG 1792, School of Business and Economics, Humboldt-Universit{\"a}t zu Berlin, Dorotheenstr. 1, 10117 Berlin, Germany; Sim Kee Boon Institute, Singapore Management University, Singapore; NUS, Center of Competitiveness, Singapore; WISE Wang Yanan Institute for Studies in Economics, Xiamen University, Xiamen, China; Yushan Scholar National Yang-Ming Chiao Tung University, Dept Information Management and Finance, Hsinchu, Taiwan, ROC; Charles University, Dept Mathematics and Physics, Prague, Czech Republic; the European Cooperation in Science \& Technology COST Action [CA19130]; Grant CAS: XDA 23020303 and DFG IRTG 1792 gratefully acknowledged}
}

\date{\today}

\begin{document}

\maketitle
\begin{abstract}
Living in the Information Age, the power of data and correct statistical analysis has never been more prevalent. Academics and practitioners require nowadays an accurate application of quantitative methods. Yet many branches are subject to a crisis of integrity, which is shown in an improper use of statistical models, $p$-hacking, HARKing, or failure to replicate results. We propose the use of a Peer-to-Peer (P2P) ecosystem based on a blockchain network, \href{https://quantinar.com/}{Quantinar}, to support quantitative analytics knowledge paired with code in the form of \href{http://www.quantlet.com/}{Quantlets} or software snippets. The integration of blockchain technology makes Quantinar a decentralized autonomous organization (DAO) that ensures fully transparent and reproducible scientific research.
\end{abstract}

\classcodes{C6, C70, I20, I21, L26, L86, O31, O36}

\pagenumbering{arabic}
\newpage

\section{Introduction}
The invention of the transistor is commonly regarded as an initial step into our modern era: The Information Age. Increased IT activity, though, naturally, brings also more data to our hands and requires proper analysis. With more computational power, this has sparked new fields like computational statistics and data science. Statisticians have quickly found that this required "proper analysis" must be based on the interplay between adequate scientific methods and efficient computational implementation.
Statistical inference and quantitative data technology have undoubtedly contributed to the scientific advances of clinical testing, financial risk management, economic forecasting, and policy making. This data-driven success story of quantitative technology has created a rich toolbox of possibly adequate instruments. The overwhelming abundance of knowledge and tools brought the necessity to develop proper analytics on top of the adequate "proper analysis".
This is where \href{https://quantinar.com/}{Quantinar} (Qr) comes into play since it offers a new peer-to-peer (P2P) system to preserve and produce knowledge, evaluate ideas, and educate the growing amount of practitioners. It offers a unified platform to perform proper data analytics since knowledge elements are sliced into "Courselets" (CLs), knowledge containers, that are created, consumed, and owned by the Quantinar community.

Quantinar uses the stable and proven concept of a platform economy. Many Web 2.0 approaches provide isolated elements, like Coursera, but do not provide horizontal and vertical integration. Another example is journals which are a proven way of retaining knowledge. 
Yet they are limited in capacity, such as the amount and speed of reviewers. They are also subject to publication bias, meaning that papers with statistically insignificant results are unlikely to be published, thus skewing research results. The existence of a publication bias of this kind is well studied \citep{Ioannidis2005}.

Education platforms offer a blend of both theoretical and practical approaches, but miss the vertical integration in terms of reproducibility, data visibility, and transparency (EdX, Youtube). They are typically top-down approaches that do not provide ownership for creators and fail to connect the creators of knowledge with their consumers and pave the way for today's consumers to be tomorrow's creators. Most importantly, those platforms cannot ensure the quality and reproducibility of research results.

We propose \href{https://quantinar.com/}{Quantinar}, a P2P platform that strengthens research collaboration and reproducibility in different areas. The architecture of Qr is extensible, and flexible and may host all kinds of scientific knowledge elements that involve quantitative analysis. Here we present the Qr system across Fintech, Blockchain, Machine Learning, Explainable AI, Data Science, Digital Economy, Cryptocurrency, and Maths \& Stats. It aims to provide better integration of scholarly articles, the studied data, and the code of the implemented analysis to ensure the reproducibility of the published results, while also providing educational content.

On top, Quantinar fosters scientific research by rewarding its participants through a modern data-driven algorithm that assesses the value of their publications. This algorithm draws inspiration from PageRank and maps the Quantinar ecosystem using a graph that represents the contributors and their contributions. The method is dynamic and flexible, ensuring temporal and spatial scalability, score inertia for each contributor, and community control.

Finally, employing a game theory-based approach, Quantinar presents a novel peer-review process that aims to incentivize community members to conduct efficient and impartial publication reviews. Through simulations, we demonstrate the efficacy of this process.

Quantinar's philosophy is Open Science and provides a framework for transparency, accessibility, sharing, and collaborative development of knowledge. Our goal is to implement Qr with Blockchain technology and smart contracts via a decentralized autonomous organization (DAO) yielding a tokenized ecosystem. With the recent developments of Web3 technology, Quantinar makes it possible to spread the benefits of AI by equalizing the opportunity to access and monetize data, code, and scientific ideas.

The first part of this paper explains the necessity of a system like \href{quantinar.com}{Quantinar} by referring to the literature and stating the current problems in modern academics. Afterward, the technology used and the status of Qr is presented. Finally, we demonstrate how Quantinar addresses some of the aforementioned issues through the utilization of cutting-edge innovations in blockchain technology and quantitative methodologies.

\section{Why Quantinar?}

According to \citet{Albright2017}, we are living in a so-called post-truth, or fake news era, where public opinion can be manipulated via different information channels. Despite the peer review process, elements of false information are also present in research publications. This may appear e.g. in the shape of publication bias \citep{Ioannidis2005} or HARKing. Thus, a methodology that improves research reproducibility and reliability is much needed in today’s academic environment and more generally in any knowledge-driven scientific system.

First, a clarification of the concept of reproducibility in scientific research is necessary. \citet{Goodman2016} provide a good definition of reproducibility by emphasizing the difference between reproducibility, replicability, and repeatability. The authors propose a new terminology:
\begin{itemize}
    \item methods reproducibility refers to the provision of enough detail about study procedures and data so the same procedures could, in theory, or actuality, be exactly repeated.
    \item results reproducibility refers to obtaining the same results from the conduct of an independent study whose procedures are as closely matched to the original experiment as possible
    \item finally, inferential reproducibility refers to the drawing of qualitatively similar conclusions from either an independent replication of a study or a reanalysis of the original study.
\end{itemize}

Inferential reproducibility might be an unattainable ideal since there might be competing models. However, the desired state of today's research, namely a framework in which authors can make arguments for or against one’s research can be designed by ensuring a transparent research process, thanks to extensive reporting of the scientific design, measurement, data, and analysis. Quantinar offers to integrate open access to scientific publications with code and data to allow for the result reproducibility and direct communication channels between researchers.

\subsection{Modern academics publish small p-values}\label{issues_academics}

Along their careers, academics and researchers face pressure to publish articles, for example: being eligible as a candidate for a Ph.D. student, being eligible for tenure at some universities, or just mere pressure from a funding private entity. This reduces the liberty of choosing the research topic and time to complete the study and can strongly impact the quality of the research process by forcing the researchers to make questionable methodological decisions that would produce "significant" results i.e. with small p-values \citep{Campbell2017}.

Analyzing the outcomes of various studies, \citet{Ioannidis2005} concludes that probably most of today's research findings are biased or even false. Indeed, Ioannidis argues that research is not most appropriately represented and summarized by p-values and research findings evaluated solely on a single study assessed by formal statistical significance, typically for a p-value less than 0.05 are most likely biased or even false. The probability that a research claim is true depends on study power, bias, and the pre-study odds, defined by Ioannidis as the ratio of true to no relationships among the relationships tested in each scientific field. 

Assessing research findings based on p-values only brings pressure on the researchers to use unethical practices such as \href{https://quantinar.com/course/35/phacking}{$p$-hacking}, or HARKing to publish their research according to what the academic environment expects of them. The term $p$-hacking refers to all the practices that could result in significant outcomes, that is having a $p$-value small enough for a null hypothesis to be rejected. Such practices could be, for example, using only a subset of the data for the estimation, choosing dependent variables post-factum, or adding data points if the final estimates are not significant \citep{Bruns2016}. HARKing is more specific and probably harder to prevent. It is defined as, "presenting a post-hoc hypothesis (i.e. based or informed on one's results) in one's research report as if it were, in fact, an a priori hypothesis" \citep{Kerr1998}. However, a modification in the experimental design after the beginning of a study may be justified for various reasons. For instance, in the medical domain, small sample sizes could be attributed to ethical considerations, whereby the drug being tested results in unexpected harm. Similarly, in Economics, unanticipated substantial costs may warrant such a change. Nevertheless, it is imperative to meticulously document these alterations in the experimental design.

While top journals are considered to act as guarantors of quality, they fail at better mitigating the problem, especially in Economics \citep{Brodeuretal:2020}. The costs of such practices span from the range of ethical issues that might arise especially in fields like medicine, to a general lack of trust in the field of science. In general, the literature seems to agree that research reproducibility practices could lower these kinds of practices.

To strengthen even further the main idea presented by Ioannidis, \citet{Fanelli2010} creates a hierarchy of scientific fields, based on the reported support for the tested hypothesis of the published papers in these fields (see e.g. p.5 in Quantinar's \href{https://quantinar.com/course/35/phacking}{$P$-hacking} courselet). The author concludes that the scientific fields that have fewer constraints for their biases (psychology, social science, etc.), usually report more positive values than the fields where these constraints are higher (space science).

\subsection{Research Reproducibility in Econometrics}

Scholarly literature in Econometrics has been long criticized for its lack of reproducibility \citep{Leamer1983}. The impossibility of sample randomization and control, which are usually not present in empirical studies, highly contributes to misplaced "true" hypotheses. Based on this critique, \citet{Ioannidis2013Economics} argue that problems like the broad flexibility of econometric models and the lack of accounting for multiplicity are problems in empirical Econometrics. To even strengthen the case, \citet{AngristPischke2010}, who brought a solid improvement in the econometrics methods, argue that Leamer's critique is only part of the problem, the other part being the selective reporting bias. However, when considering those issues, the conclusion is that "strengthening the reproducibility culture with an emphasis on independent replication, conducting larger, better studies, promoting collaborative efforts rather than siloed [...], and reducing biases and conflicts" are ways in which contemporary research in Econometrics can be improved.

On top, since empirical studies are hard to realize in Econometrics, researchers strongly rely on data. In particular, data-hungry methods such as machine learning and non-parametric statistics are used more frequently and cannot be reproduced since the data is often private because of the cost of building, curating, maintaining, storing, etc. \href{http://www.quantlet.com/}{\quantlet Quantlet} \citep{Hardle2007Quantnet, Borke2017}, the \href{https://blockchain-research-center.com}{Blockchain Research Center} or \href{https://paperswithcode.com/}{PapersWithCode} are platforms that already promote data and code accessibility as open access libraries. On top, solutions such as \href{https://www.cascad.tech/}{CASCaD}, \href{https://www.replicabilitystamp.org/}{ReplicabilityStamp} and \href{https://codeocean.com/}{Code Ocean} are trying to address the reproducibility problem by providing a reproducibility stamp to scholarly publications. However, a larger integration between data, code, information, and researchers is necessary.

\subsection{Can the peer review process ensure quality?}

Finally, the last problem Quantinar tries to address concerns the peer review process. The peer review process should ensure the quality of the research produced by top journals, however, its necessity and effect on scientific research have been discussed and criticized for decades \citep{Ziman1968, Spier:2002, Rowland:2002, Ware:2011}. The goal of this paragraph is not to give another extensive review of it but to give a summary of the literature. \citet{Ross-Hellauer:2017} gives a good definition of the generic peer review process as the formal quality assurance mechanism whereby scholarly manuscripts (e.g. journal articles, books, grant applications, and conference papers) are made subject to the scrutiny of others, whose feedback and judgments are then used to improve works and make final decisions regarding selection (for publication, grant allocation or speaking time). Its function is twofold: evaluating the validity and assessing the innovation and impact of the submitted work.

The peer review process is used across various disciplines and it is widely agreed that it participates in maintaining the overall quality of the scholarly literature \citep{Rowland:2002} in particular in medical sciences \citep{Lock:1985}. Nevertheless, multiple surveys have identified a widespread belief that the current model is sub-optimal \citep{ALPSP:1999, Ware:2008} which is caused by the multiple critics against the peer review process. \citet{Ross-Hellauer:2017} distinguishes 6 categories among those critics:
\begin{itemize}
    \item the unreliability and inconsistency of the reviews which is inherent to human judgment
    \item the delay between submission and publication which slows down the research progress and the expenses associated with the peer review process
    \item the lack of accountability of the involved agents (authors, editors, and reviewers) and risks of subversion introduced by the anonymity of the reviewers (in particular for single blinded reviews)
    \item social (gender, nationality, institutional affiliation, language, and discipline) and publication biases (preference of complexity over simplicity, conservatism against innovative methods, preference for positive results against negative or neutral ones which lead to p-Hacking or HARKing defined in Section \ref{issues_academics})
    \item the lack of incentives for reviewers
    \item and finally the wastefulness: the "black-box" nature of the process hides discussions between reviewers and authors that could benefit younger and future researchers.
\end{itemize}

Moreover, \citet{Heckmanetal:2020} show that the top journals in Economics fail to ensure a higher quality of their publications. By comparing cumulative citation counts (measured as of 2018) of articles published in the top5 and those published in 25 non-top5 journals over the ten-year period 2000–2010, the authors argue that non-top5 journals can produce as many if not more influential articles than the top5, concluding that, whether an article is published in the top5 or not, is a poor predictor of the article’s actual quality in the Economics literature. Finally, \citet{Ellison:2011} argues that with the development of the Internet, the necessity of peer review has lessened for high-status authors by observing a decline of publications from Economists in top-ranked university departments between the early 1990s and 2000s.

Indeed, with the development of the internet and in response to those critics, multiple solutions have been suggested for disseminating research as part of the Open Science movement and in opposition to the traditional scientific publication process. \citet{vicente:2018} identifies the core elements of Open science, which are the transparency, accessibility, sharing, and collaborative development of knowledge, and defines Open science as "transparent and accessible knowledge that is shared and developed through collaborative networks". Platforms such as \href{https://arxiv.org/}{arXiv} or \href{https://www.ssrn.com/}{SSRN} gather pre-print versions of scientific articles. \href{https://www.academia.edu/}{Academia} or \href{https://www.researchgate.net/}{ResearchGate} are community-based platforms that act as social media for researchers where they can upload their articles to gain visibility, connect with other researchers to communicate and increase their network. On top, researchers can easily share their code on \href{https://cran.r-project.org/}{CRAN} network as R packages or simply as a repository on \href{https://gitlab.com/}{GitLab} or \href{https://github.com/}{GitHub} in any language, which considerably help the associated research reproducibility. Finally, communities such as \href{https://huggingface.co/}{Hugging Face} or \href{https://www.kaggle.com/}{Kaggle} allow users to collaborate on common tasks by sharing pre-trained models, code, or datasets in the machine learning universe.

Nevertheless, the previously cited platforms have a centralized infrastructure usually controlled by a private institution whose interests do not necessarily align with the community they represent. On top, the data storage is often centralized via a cloud solution such as AWS or Google Cloud and the institution can revoke that access at any time. Finally, Open science cannot free itself from the need for quality control of information via the peer review process. While the literature provides multiple proposals around Open peer review \citep{Ross-Hellauer:2017} (see section \ref{sec:opr}) and nowadays, many scholarly journals employ versions of open peer review practice, including BMJ, BMC, Royal Society Open Science,  Nature Communications, the \citeauthor{PLOS} journal, no solution engages a scientific community in an open manner where the researcher and his/her ideas are at the center.

As we have outlined in the previous sections, independent solutions addressing specific issues exist. Our goal is to integrate those ideas into a single platform, namely Quantinar, that should set a new standard for scholarly publications thanks to vertical integration of article, code, and data to allow and verify research reproducibility, a fair and transparent open peer review process, full control for the researcher over his publications.

\section{What is Quantinar?}

\href{https://quantinar.com/}{Quantinar} is a peer-to-peer (p2p) platform that strengthens research collaboration and reproducibility in different areas like Fintech, Blockchain, Machine Learning, Explainable AI, Data Science, Digital Economy, Cryptocurrency and Maths \& Stats. It aims to provide better integration of scholarly articles, the studied data, and the code of the implemented analysis to ensure the reproducibility of the published results. The conceptual architecture of Quantinar is called \textit{C5} and it stands for \textit{Creation, Content, Consumption, Coins, and Chain}. Blockchain technology (\textit{chain}) and smart contracts via a decentralized autonomous organization (DAO) with a tokenized ecosystem (\textit{coins}) can be used today to promote Open Science, transparency, and accessibility of research and collaborative development of knowledge \citep{tenorioetal:2019}. In this section, we will present the features, applications, infrastructure, and technology of Quantinar.

Quantinar is organized as a Decentralized Autonomous Organization (DAO), an online community that jointly controls the organization's funds to pursue common goals
(\cite{Buterin2013}, \cite{Buterin2014}). As described in \ref{sec:reputation}, the Quantinar DAO will represent a P2P platform for knowledge sharing that will have a reputation-based governance, meaning that only active contributors in the academic environment will be able to control the update of the platform and the management of its funds. Quantinar's main goal is to enforce scientific research reproducibility by creating a new digital publication platform that requires more input from scholars, namely giving access to the code, data, results, and extra information provided by courselets (see Section \ref{sec:content} for a definition of such a knowledge unit). Moreover, the peer-review process described in \ref{sec:opr} ensures the quality control needed by an academic journal. On top of that, Quantinar is designed to become an integrated solution for presentation, data loading and exploration, code execution, and computing power. By creating an integrated research environment, the platform will also offer a marketplace for Datalets (Dl), Models, or Quantlets (Ql) (see Section \ref{sec:content} and \ref{sec:qr_ecosystem}) that aims to connect the academic environment to the industry. This will generate the revenue needed for the growth of the platform, and for other causes like funding research that the community decides it's needed. The governance process and tokenomics research of Quantinar will be described in a separate publication. At the time of writing, Quantinar (\href{https://www.quantinar.com/}{quantinar.com}) is a website that manages the C5 paradigm.

\subsection{Content and Creation}\label{sec:content}

The main content on Quantinar is a "CourseLet", abbreviated as CL. A CL is a scientific knowledge unit, such as a research study or course unit, in the form of slides with a presentation video and the associated PDF file of the slides for reading. If applicable, it must be accompanied by a Quantlet and a Datalet, the associated implementation, and data. Any user of Quantinar can create a courselet that will have three statuses: unverified, verified, and peer-reviewed. The unverified status is the default status obtained at the time of publication on Quantinar. The verified status is given if the courselet contains the slides with the link to the verified QuantLet (QL) and DataLet (DL), ensuring that the research results can be reproduced. The peer-reviewed status can only be obtained after passing the open peer review process defined in Section \ref{sec:opr}. Finally, a simplified pre-publication review will be used to verify that the uploaded content complies with the platform's rules, using e.g. \includegraphics[scale=0.05]{figures/quantlet.png} \href{https://github.com/QuantLet/AOBDL_code}{AOBDL} to avoid offensive content.

The choice of the CL video format is not insignificant. Indeed, videos allow a faster sharing of research studies compared to scholarly articles and emphasize scientific ideas and results generation, not formal writing. On top, a video does not replace scholarly articles that can always be referred to in the CL but rather motivates reaction and direct communication between Quantinar community members via our discussion channels (see Section \ref{sec:community}).

Finally, the authors keep their copyright and full ownership of their creation as each CL in the C5 framework is mapped to a Non-Fungible Token (NFT), a coin in the blockchain (see Section \ref{sec:nft}) and they give an open access license instead.

\subsection{Features}

Quantinar's features are designed to benefit the scientific community as a whole. The main planned or already implemented features are summarized in Table \ref{table:feature}. In particular, Quantinar answers the needs of three main categories of users. By offering open access to CLs and certification, students and professionals can acquire skills necessary for their careers. By offering a classroom and the possibility to reuse other users' courselets, universities and professors can attract more students and gain online visibility. Finally, researchers increase their reputation by creating transparent and reproducible research, reviewing other users' publications, and share knowledge via a simplified publication and communication channel.

\begin{table}[!t]
\caption{Quantinar's features\label{table:feature}}
\begin{tabular*}{\columnwidth}{@{\extracolsep\fill}lll@{\extracolsep\fill}}
\toprule
Feature's Type & Feature & Status \\
\midrule
    \multirow{6}{*}{Content creation } & Upload Courselet & Done \\
    & Link QuantLet & Backlog \\
    & Link DataLet & Backlog \\
    & Make a course from multiple CLs & Done \\
    & Make a classroom & Backlog \\
    & Start a blog & Backlog \\
    & & \\
    \multirow{4}{*}{Content consumption } & Explore Courselets & Done \\
    & Explore QuantLets & Backlog \\
    & Explore DataLets & Backlog \\
    & Read Courselet slides & Done \\
    & Watch Courselet video & Done \\
    & Obtain a course certificate & In Progress \\
    & Experiment with code and data & Backlog \\
    & Read blogs & Done\\
    & & \\
    \multirow{3}{*}{Community} & Comment Courselet & Done \\
    & Review Courselet & Done \\
    & Likes and other reaction & Backlog \\
    & Discuss on Discord & In progress \\
    & ... & \\
    & & \\
    \multirow{2}{*}{Open peer review} & Call for reviews &  Backlog \\
    & On chain review process & Backlog \\
    & ... &  \\
    & & \\
    \multirow{2}{*}{Blockchain} & DAO &  Backlog \\
    & Governance & Backlog \\
    & ... &  \\
    & & \\
\bottomrule
\end{tabular*}
\end{table}

\subsection{Applications}\label{sec:applications}

Quantinar's purpose is to become a platform that can be used by any person that works in the field of data analysis, analytics, and science, namely \textit{students, teachers, and researchers}.

It can be seen as a \textit{student platform} because students can easily understand individual research problems in the form of courselets that offer slides and presentation videos accompanied by code examples that are already implemented by top researchers. Moreover, students can enroll in courses that tackle a wider range of subjects and get a certification after passing a series of tests to verify their knowledge. Running algorithms in the cloud, exploring datasets and models, and sharing their work will all be integrated under the same platform that makes learning and acquiring knowledge interactive and accessible.

Quantinar is a \textit{teacher platform} as much as it is a student one. Teachers can create CLs $\{$flowers$\}$ which link other CLs to their courses, to compose more complex and more complete teaching environments. Apart from that, teachers can also get coins, reputation, and citations when others use their courselets, a feature which is thoroughly described in \ref{sec:reputation}

For \textit{researchers}, Quantinar delivers the ability to publish CLs that represent their research articles and projects. By doing so, they gain coins (QNAR and QLET), reputation, and citations when other people use their research (see Section \ref{sec:reputation}). Quality is ensured by a P2P review process that is later described in \ref{sec:opr}. By publishing their work in the form of courselets, researchers must publish a video presentation, code, and data (when possible), on top of the article itself, which strengthens the reproducibility of the results.

\subsection{Prosumers and consumption}
\begin{figure}[!ht]
    \centering
    \includegraphics[width=0.6\textwidth, keepaspectratio]{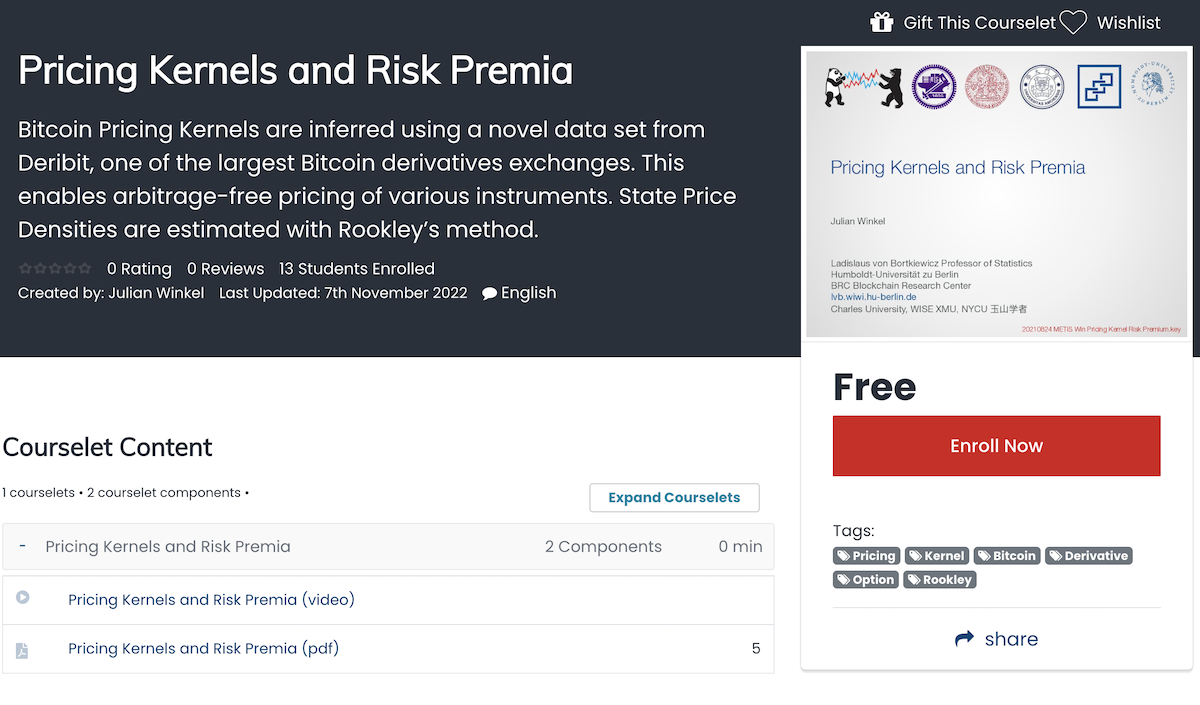}
    \caption{Courselet: Pricing Kernels}
    \label{img:courselet-pricing-kernelts}
\end{figure}

Quantinar is also a prosumer platform. This means that students can become teachers by creating CLs of their own. Researchers are also encouraged to post their preliminary results and get feedback from the community while also contributing to its growth. In \autoref{img:courselet-pricing-kernelts} we can find an example of a CL, that represents work done by one of the authors of this paper.

\subsection{Case Study: Extending a course}
\begin{figure}[!ht]
    \centering
    \includegraphics[width=0.6\textwidth, keepaspectratio]{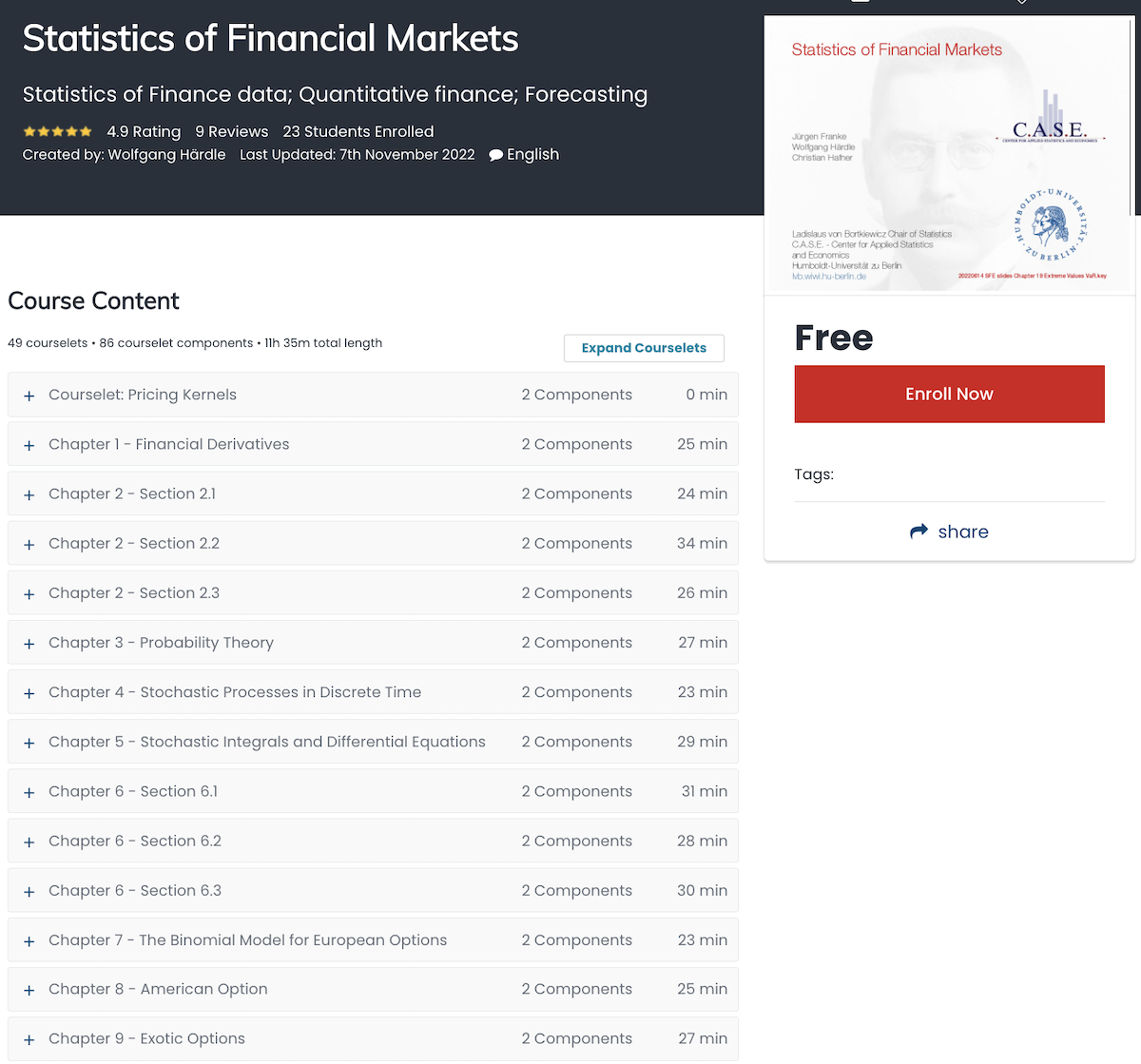}
    \caption{Course: Digital Economy and Decision Analytics}
    \label{img:course-deda}
\end{figure}

As we briefly mentioned in Section \ref{sec:applications}, Quantinar is also a platform for teachers. Teachers can compose courses using both their content and CLs that are already posted on the platform by other prosumers. As we can see in Figure \ref{img:course-deda}, the CL created in Figure \ref{img:courselet-pricing-kernelts} was made available in the \href{https://quantinar.com/course/103/statistics-of-financial-markets}{Statistics of Financial Markets (SFM) Course}. Of course, the other courselets that are available in the SFM course can be further linked to develop courses that are more suitable for other teacher's visions

\subsection{Quantinar in numbers}\label{sec:community}

The Quantinar ecosystem already gathered an international community of researchers that produces quality research on diverse scientific topics. At the time of writing, Quantinar already has an important community and can offer high-quality content with 215 CLs.

\begin{figure}[h!]
	\centering
	\includegraphics[width=0.6\textwidth, keepaspectratio]{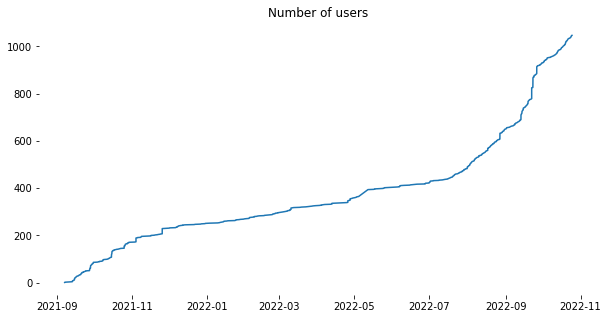}
	\caption{Number of users}
	\label{fig:users_ts}
\end{figure}

Quantinar’s community consists of 1470 users with 43 instructors. Without a marketing campaign, Quantinar has experienced important growth in the last months as it is shown in Figure \ref{fig:users_ts}.

\begin{figure}[h!]
	\centering
	\includegraphics[width=0.6\textwidth,]{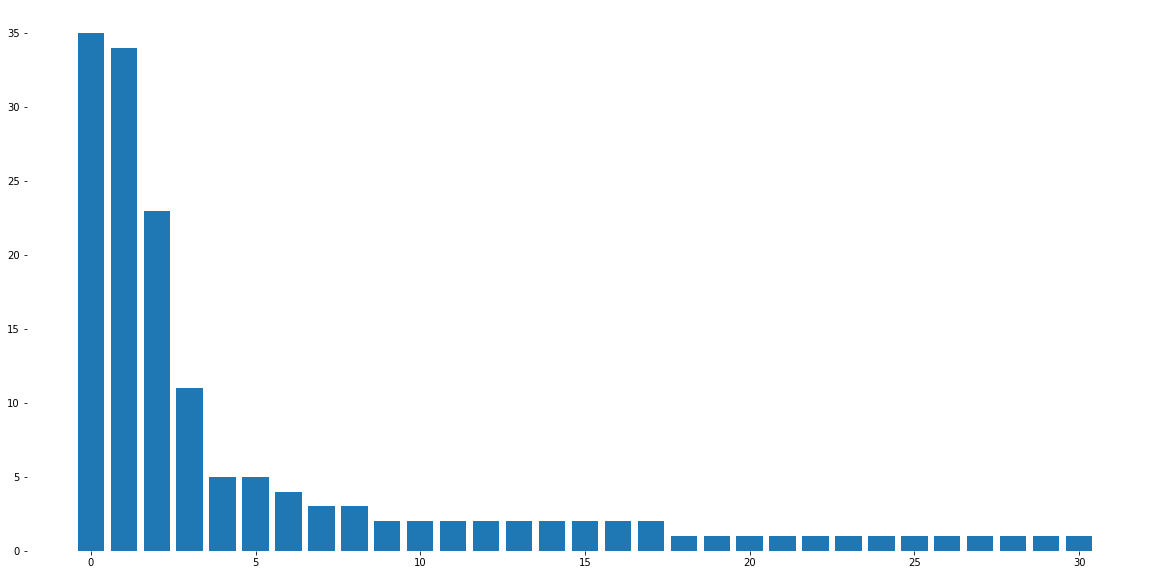}
	\caption{Number of courselets per instructor}
	\label{fig:courselet_users}
\end{figure}

At the time of writing, the distribution of CL instructors contributing to the CL creation is slightly skewed (Figure \ref{fig:courselet_users}).

\begin{figure}[h!]
	\centering
	\begin{subfigure}[b]{0.45\textwidth}
		\centering
		\includegraphics[width=\textwidth]{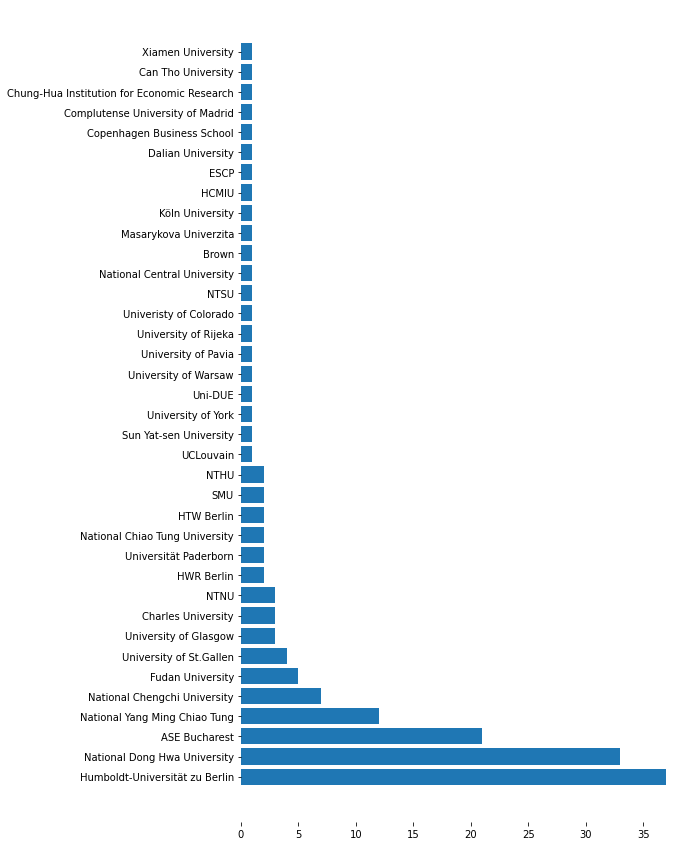}
		\caption{Users' affiliation}
		\label{fig:users_affiliation}
	\end{subfigure}
	\hfill
	\begin{subfigure}[b]{0.45\textwidth}
		\centering
		\includegraphics[width=\textwidth]{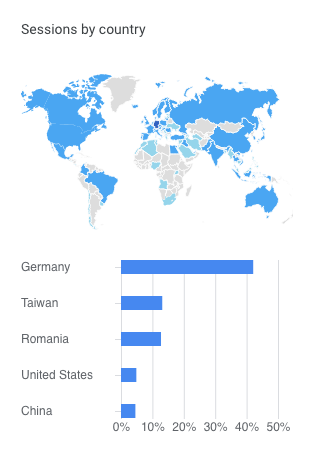}
		\caption{Sessions by country}
		\label{fig:sessionslocation}
	\end{subfigure}
	\caption{Users location}
	\label{fig:users_location}
\end{figure}

This is an effect of where the original ideas come from: \href{https://www.wiwi.hu-berlin.de/en/}{Humboldt-Universität zu Berlin} and \href{https://www.ase.ro/index_en.asp}{ASE Bucuresti}. The Quantinar ecosystem has already attracted users from multiple countries and continents (Germany, Romania, United States, China, Singapore, Taiwan, etc.), as it is shown in Figure \ref{fig:users_location}.

Thanks to its Discord channel (\href{https://discord.gg/ebS3Bf6gfS}{https://discord.gg/ebS3Bf6gfS}), Quantinar's community is thriving: members can follow the latest updates from the platform and interact with it.

\begin{figure}[h!]
	\centering
	\includegraphics[width=0.5\columnwidth, keepaspectratio]{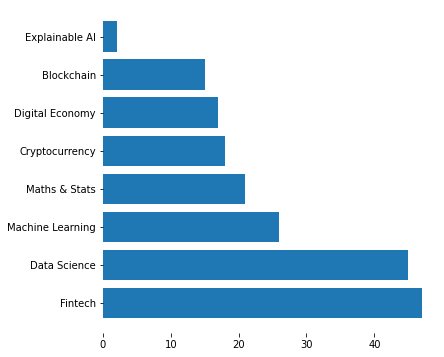}
	\caption{Courselet category distribution}
	\label{fig:courseletcategories}
\end{figure}

\begin{figure}[h!]
    \centering
    \begin{subfigure}{0.6\textwidth}
        \centering
        \includegraphics[height=4cm, keepaspectratio]{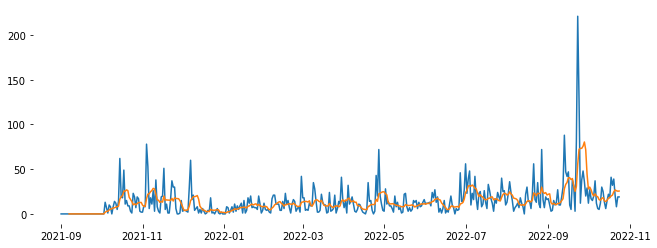}
        \caption{\textcolor{blue}{Total daily unique views} on Qr's courselets and \textcolor{orange}{week rolling average}}
        \label{fig:ts_views}
    \end{subfigure}
    \bigskip
    \begin{subfigure}{0.6\textwidth}
        \centering
        \includegraphics[height=4cm, keepaspectratio]{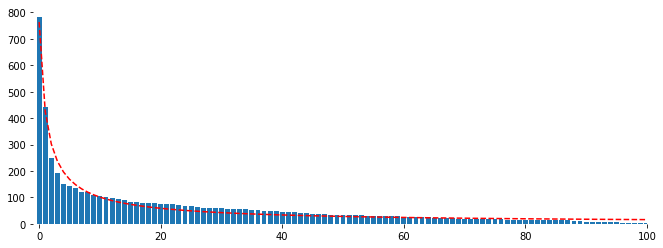}
        \caption{Total unique views per CLs}
        \label{fig:views_hist}
    \end{subfigure}
    \caption{Qr's courselets views}
\end{figure}

At the time of writing, Quantinar gathers 245 courselets and 5 courses in 8 categories (see Figure \ref{fig:courseletcategories}). 
We can observe in Figure \ref{fig:ts_views} an uptrend in the consumption of courselets on Qr as, in the second half of 2022, there are around 21 unique CL views per day, while there were only 9 before.  The total views per courselet, $y$, follows a classic power law given by $y = ax^k + b$ for $x >0$ with parameters $(k, a, b) = (763, -8.45,  0.39)$. The top 3 CLs are \href{https://quantinar.com/course/23/Blockchain}{DEDA Digital Economy \& Decision Analytics}, \href{https://quantinar.com/course/103/statistics-of-financial-markets}{Statistics of Financial Markets} and \href{https://quantinar.com/course/67/ADM}{Advanced Mathematics} gathering more than 25\% of all courselets views (see Figure \ref{fig:views_hist}) and the top 20\% CLs corresponds to 60\% all views. 

\begin{figure}[h!]
	\centering
	\includegraphics[width=0.8\columnwidth, keepaspectratio]{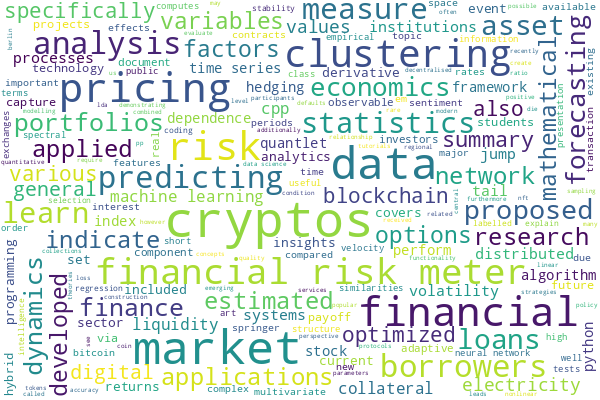}
	\caption{Word occurrence in courselet description}
	\label{fig:coursedetailwordcloud}
\end{figure}

\begin{figure}[h!]
	\centering
	\includegraphics[width=0.6\textwidth, keepaspectratio]{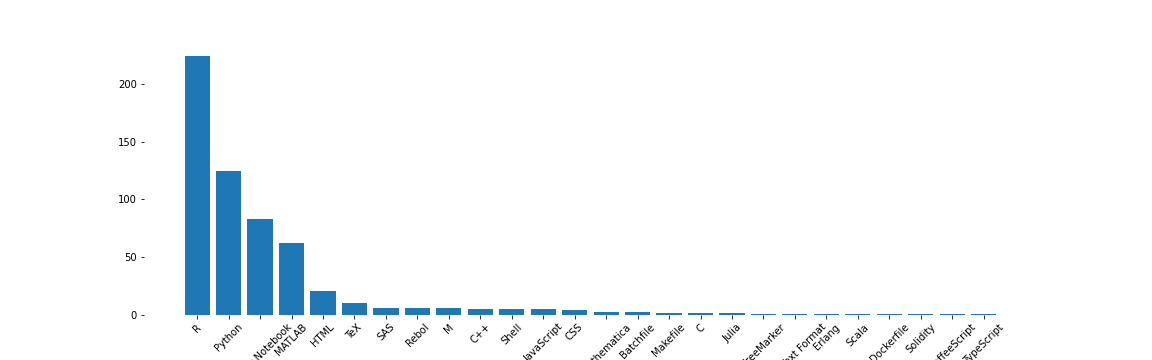}
	\caption{Quantlets programming language}
	\label{fig:quantletrepolang}
\end{figure}

The most recurrent topics are related to cryptos, clustering, price, prediction, marketing, or finance, as the word cloud in Figure \ref{fig:coursedetailwordcloud} suggests it.  Finally, Quantinar's CLs come with their respective implementation code in \href{http://www.quantlet.com/}{\quantlet Quantlet} using mostly R or Python programming languages (see Figure \ref{fig:quantletrepolang}). 

\subsection{The Quantinar Ecosystem}\label{sec:qr_ecosystem}
The concept of the Quantinar ecosystem refers to achieving smooth and rich integration between the DL components \href{https://blockchain-research-center.com}{Blockchain Research Center} (BRC), the \href{https://quantlet.com}{Quantlet} code base and the \href{https://quantinar.com}{Quantinar} platform. A possible goal of this integration is to establish interactive code environments generated from the (QLs) that are present in the CLs in Quantinar. This architecture ensures that the CL results and claims are easily reproducible \citep{Borke2017Dynamic} as e.g. this current paper at \quantlet \href{https://github.com/QuantLet/quantinar-rep}{Quantinar-rep} with respective \href{https://www.quantinar.com/course/525/quantinar-a-p2p-knowledge-platform}{courselet}, \citep{Spilak2021} at \quantlet \href{https://github.com/QuantLet/MLvsGARCH}{MLvsGARCH} or \citep{spilak2022does} with associated \href{https://www.quantinar.com/course/111/does-non-linear-factorisation-of-financial-returns-help-build-better-and-stabler-portfolio}{courselet} and QL \quantlet \href{https://github.com/QuantLet/EmbeddingPortfolio}{EmbeddingPortfolio}. Moreover, the runtime execution environments will be able to access the data that is present in the DL components, e.g. the BRC.

\begin{figure}[h!]
	\centering
	\includegraphics[width=0.6\textwidth, keepaspectratio]{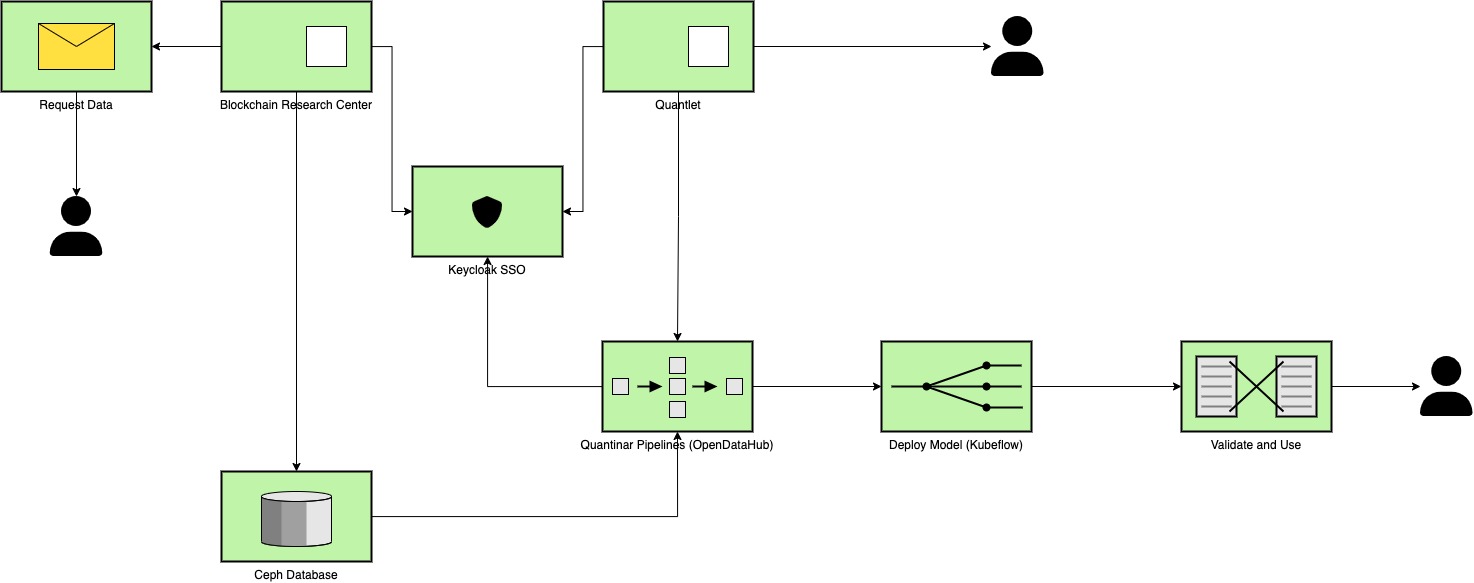}
	\caption{Quantinar Ecosystem Integration Diagram}
	\label{fig:quantinar-ecosystem}
\end{figure}

This is achieved by moving the DLs from the BRC to a Ceph (\href{https://ceph.io/en/}{ceph.io} storage backend, which is integrable into a Kubernetes (\href{https://kubernetes.io}{kubernetes.io}) cluster, that runs OpenDataHub (\href{https://opendatahub.io}{opendatahub.io}). The high-level architecture overview can be seen in Figure \ref{fig:quantinar-ecosystem}.

\section{How? Blockchain, a technology for Open Science}

\subsection{Technology}\label{sec:technology}

\begin{figure}[h!]
    \centering
    \includegraphics[width=0.6\textwidth, keepaspectratio]{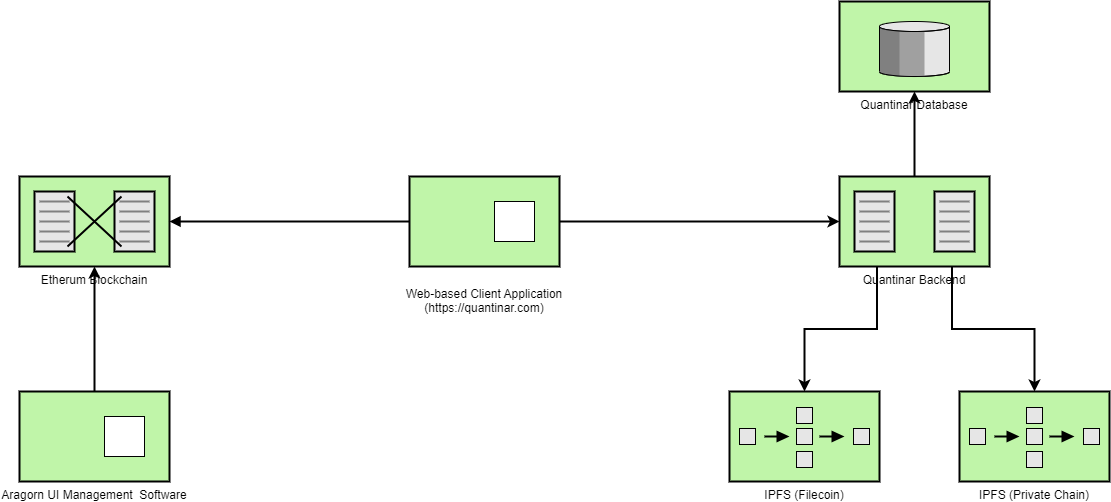}
    \caption{Software Architecture Overview}
    \label{fig:software-architecture}
\end{figure}

The Quantinar P2P platform - abbreviated Qr - will have most of its operations on-chain since having decentralization in its core is a strong suit compared to most educational platforms. However, as for academic journals, Quantinar must ensure that each user is indeed a real person and cannot transfer or sell its wallet addresses, and with them, its acquired reputation. To do so, a centralized software component with specific capabilities is required, as illustrated in Figure \ref{fig:software-architecture}.

Indeed, as accountability in the OPR process cannot be achieved without an immutable online identity, the centralized component should be able to provide identity management features such as a Single Sign On or BrowserID solution \citep{sso}.

That way, Quantinar can integrate all the different components of the platform under the same umbrella. To do so, the Keycloak project \citep{keycloak} is used, with a simple MySQL Database. An extra layer of information will be sought from the users thanks to a KYC solution, for example, MouseKYC \citep{mousekyc}.

Except for identity management, a minimal gateway is needed to provide flawless document uploading to an IPFS chain and validate the uploaded content to filter out forbidden items or spam. Articles are uploaded on a public IPFS chain like Filecoin's \citep{filecoinWhitepaper}, while presentations, videos, and other types of content will be uploaded on a private IPFS chain managed by the Quantinar DAO. More information about IPFS can be found under section \ref{sec:ipfs}.

The Quantinar DAO will be hosted on the Ethereum blockchain, as it is one of the most stable and predictable technology. Moreover, the recent adoption of Proof of Stake (POS) and the upcoming implementation of sharding by Ethereum improves its competitiveness concerning other blockchains. Indeed, sharding, defined as the process which divides one blockchain network into several smaller networks, referred to as "shards", improves the performance of blockchains that implement it by giving them a competitive advantage in terms of speed for transaction processing and data availability. On top, the planned launch of SoulBound tokens, announced in early 2022 by Vitalik Buterin (\href{https://vitalik.ca/general/2022/01/26/soulbound.html}{vitalik.ca/soulbound}) comes in good time. SoulBound tokens are \textit{special NFTs with immutable ownership and pre-determined immutable burn authorization}. Quantinar is a perfect application for SoulBound tokens via the various certifications obtained when passing a course.

Finally, the whole Web3 side of the platform is written using the Aragon framework (\href{https://aragon.org/}{aragon.org}), one of the most popular DAO frameworks on Ethereum. The centralized side of the infrastructure will be hosted on an OpenStack cloud fully installed and managed by the Quantinar Team.

\subsection{Open Access and IPFS}\label{sec:ipfs}

At the center of Quantinar is Open Access to its content. Nevertheless, it is not enough to guarantee Open Access to content if it is controlled by a centralized institution such as an online academic journal. Indeed, the access can be revoked at any time, or even worse, ransomware attacks could take place which would render all of the data on the servers useless. 

The InterPlatenary File System (IPFS) is a peer-to-peer protocol that distributes data storage in its network in a decentralized manner, firstly described in "IPFS - Content Addressed, Versioned, P2P File System" as \textit{an ambitious vision of new decentralized Internet infrastructure, upon which many different kinds of applications can be built} \citep{benet:2014}.  It builds on top of common ideas collected from pieces of software like BitTorrent or Git, to create a protocol that manages Merkle DAGs containing file data over a network.

There have been multiple attempts at building an IPFS-based storage solution that is ready for solving real-world challenges like decentralizing medical data, as described by \citep{kumar:2021}, and even more attempts of building IPFS networks for academic usages, like the share of databases used for research \citep{meng:2021}, some of which tackled issues like availability, immutability, transparency, and security \citep{kumar:2021}, and even open access to academic research \citep{kumar:2020}.

Decentralization makes it difficult to restrict the network's data access. On top of this, thanks to content addressing, files stored using IPFS are automatically versioned (https://ipfs.io/). Thus, by using a decentralized storage system such as IPFS, one cannot revoke access to the authors' publications or delete an open peer-review article, as Quantinar's content is shared across an IPFS network. Finally, at times of writing, Filecoin costs are extremely low compared to standard centralized cloud-based solutions (<1\$/month/TB and 23\$ for Filecoin and AWS respectively, source \href{https://largedata.filecoin.io/}{Filecoin} and \href{https://aws.amazon.com/s3/pricing}{AWS}). Quantinar's network and reward system is then truly resilient.

For the moment, there is no plan to implement a proprietary IPFS network. This is unnecessary since the papers should be stored on public chains, so they can be openly accessed. On top of that, a scalable and resilient IPFS network is required for the platform to take off. For these reasons, we propose the usage of the Filecoin network (\citep{filecoinWhitepaper}). The capabilities of token generation of the Quantinar DAO can help support the filecoin storage on a \textit{pay as you grow} basis.

Of course, private data use cases will be developed in a future phase of the platform that would require a private IPFS network with access control (as seen in \citep{steichen:2018}). This will be designed specifically for holding private databases, original LaTeX files, or other information which might be seen as sensitive in the eyes of the researchers, companies, or universities. Thus, all of the actors will have the power to decide who can access this information. More research and development are required for developing such a feature, which is planned to happen in the second phase of the project.

\subsection{Ownership and copyrights: Courselet NFT}\label{sec:nft}

Quantinar makes sure that the researchers keep the copyrights and ownership of their publication, CLs, DLs, or QLs. For that goal, Quantinar offers to create a non-fungible token (NFT) for each CL on the platform and transfer it to the CL author(s). An NFT is a unique cryptographic token created from a specific smart contract standard (e.g. ERC-721 on Ethereum blockchain) that provides functionalities such as ownership transfer or ownership verification (see \citep{doi:10.1142/9781800612723_0001} for an extensive study of smart contracts applications). Thanks to this NFT, any authors can claim ownership of a specific courselet on the Quantinar platform by providing the unique hash of the NFT associated with it.

While smart contract and NFT makes it easy to verify ownership of a digital object on the blockchain, copyrights exist in the non-digital world and are governed by state law. Thus, to truly ensure copyright protection, each courselet will be secured with an Open Source license such as MIT or GNU. Finally, authors will keep ownership of the intellectual property (IP) on their work allowing them to publish their study in external academic journals.

Some information such as authors' identity, CL title, link to the CL content, review score, and link to reviews will be stored on-chain, while to reduce storage cost the CL content and actual reviews will be stored off-chain on Quantinar IPFS network defined in the previous section \ref{sec:ipfs}. Having a courselet NFT with an immutable link to versioned content and reviews helps to open the iterative publication process during the peer-review feedback loop and protect the integrity of the authors' reputation score defined in \ref{sec:rep_score}. Thus, on top of the CL metadata, all contribution logs will be accessible directly on the blockchain ledger or on the IPFS network ensuring transparency.

\subsection{Contributor reward}\label{sec:reputation}

In the context of academic publications, the Bibliometrics research field has a long tradition and proposed many metrics for measuring the global productivity and impact of research in science and society. The most famous metric is probably the H-index \citep{doi:10.1073/pnas.0507655102} which is an author-specific measure. It can be highly relevant in a scholar's career as it is often taken into consideration for allocating resources between researchers and funding. Because of many critics, mostly related to the relative easiness of the H-index manipulation \citep{Bartneck:2011}, researchers proposed alternative measures, in particular altmetrics which estimates the value of a publication at the article level based on various web data sources publicly available, e.g. social media such as Twitter \citep{altmetric}. Those metrics are highly valuable for journals and publishers and are often directly integrated into their publications. However,  both bibliometrics and altmetrics remain weak indicators as they can easily be gamed \citep{Karanatsiou:2017}. An interesting comparison between academic ranking methods in Economics is given by \citet{Zharovaetal2023} who analyses the use of resources and scientific performance on the levels of individuals, research groups, and universities.

Quantinar offers a valuation system that measures the relative value of a contribution, or reputation, within the Qr ecosystem. The Qr reputation system is scalable in time with the size of the graph and the score reflects the inertia of highly valuable contributions. Indeed most of the results today can be attributed to the labor provided yesterday and contributions that are highly valuable should be in the long run. The inertia of the reputation score of some contributor $i$, $S_i^t$ over the periods $t$ can be measured by the rate of change of the score assuming no change between the successive periods. Finally, the system must make sure that evaluations cannot be gamed in favor of malicious agents to be sustainable.

Since the development of online P2P communities, multiple proposals have been made to ensure trust between members and the quality of members' contributions to develop sustainable P2P platforms. A well-known technique is the PageRank algorithm which the scaling factor is roughly linear in $\log(n)$ \citep{page1999}. While it was originally built for ranking web pages in the context of improving search engines, it serves also as an indicator of the trustworthiness of a website. Since its publication, it has been widely studied (the reader can refer to \citep{6998874} for a short survey or \citep{LangvilleMeyer+2011}) and extended for example, EigenTrust \citep{10.1145/775152.775242} evaluates trust in a distributed-manner within P2P file-sharing communities. Some alternatives have been proposed such as PeerTrust \citep{xiong:2004} which evaluates the members' reputation based on specific contextualized parameters such as contribution feedback, number of contributions, and credibility of the feedback source. However, PageRank is not famous only because it was created by the founders of Google, but its simplicity, generality, guaranteed existence, uniqueness, and fast computation are the reasons that it is used in many other applications than Google's search engine and is still very popular today. In fact, \citet{doi:10.1137/140976649} shows how to apply PageRank to biology, chemistry, ecology, neuroscience, physics, sports, and computer systems. Thanks to PageRank, we can estimate the value of any contribution relative to the marginal value it brings to the community as a whole and engage the community by effectively rewarding Quantinar contributors for the labor they provided to produce their contribution.

To estimate the reputation of each member, we propose to use a modified version of PageRank named CredRank \citep{credrank}, an algorithm developed for blockchain-based communities by SourceCred (\href{sourcecred.io}{sourcecred.io}). The innovation of CredRank is to embed temporal information into the graph on which PageRank is computed. As we will explain in the next section, thanks to CredRank, only one PageRank invocation is necessary to obtain the reputation score defined by \eqref{eq:rep_score} for each Qr member at any period. This property of CredRank greatly improves the scalability of the reputation score evaluation over time. The inertia of past contributions can easily be controlled without having to run PageRank again and each period's importance can easily be evaluated.

SourceCred provides an algorithm for distributing rewards within a community that is sufficiently abstract to be used in any DAO thanks to the concepts of \textit{Grain} and \textit{Cred}. Cred refers to the reputation score defined in the next Section \ref{sec:rep_score} which is mapped by a utility token that is not transferable and can be gained by contributing to the community. In the Quantinar DAO, the Cred token will be named QNAR. The second concept of SourceCred's algorithm is Grain, represented by the Quantlet token, QLET, in Quantinar DAO. This token is meant for the creation of an internal research-based economy and will be tradable on external exchanges. It will be used to pay contributors based on the QNAR score and gain access to premium features of the platform such as private courselet components or consulting top-level researchers (see Section \ref{sec:reward}).

\subsubsection{Reputation score evaluation}\label{sec:rep_score}

The first step for computing the QNAR-based reward a Qr user should receive for its contribution is to evaluate its reputation within the community thanks to the PageRank algorithm. Therefore the Qr ecosystem is mapped onto a graph with specific nodes of three types: the seed node $s_0$, contributors, and contributions forming the sets $\mathcal{C}$ and $\mathcal{A}$, respectively. The seed node mints the new QNAR tokens and distributes them proportionally according to the contributions in the graph. Finally, the minted tokens flow from the contributions to the contributors who eventually receive a part of the originally minted tokens relative to their participation in Quantinar’s total value. The following non-exhaustive list of contributions will be mapped onto the graph: creating a CL, QL, or DL, viewing, citing, and reviewing a CL, enrolling as a student into a CL, creating a course linking other CLs, actively participating in the discord channels, voting in the DAO decision process and helping other community members with their research. It is already clear that not all contributions should be rewarded equally, as creating a CL requires much more labor than citing one. Thus weights are associated with each contribution type in a heuristic manner based on what the community values in Qr and determine how much QNAR is minted for each contribution (see Table \ref{table:nodes_weights}). SourceCred finds the equilibrium of the token distribution among community members, in other words, it finds the stationary distribution of the PageRank random walk associated with the community graph. Thanks to the contribution weights, the Qr community keeps full control of the distribution of the reward among contributions.

In detail, the source node $s_0$, contributions, and community members are mapped onto a directed graph $G = (\mathcal{V}, \mathcal{E}, \mathcal{W})$ where $\mathcal{V} = \{s_0\} \cup \mathcal{A} \cup \mathcal{C} = \{v_i\}_{1\leq i \leq n}$ is the set of $n$ unique vertices mapping the seed node, contributions and contributors which are connected with $m$ directed edges, $\mathcal{E} = \{e_{ij}\}_{(v_i,v_j) \in \mathcal{V}^2}$, where $e_{ij}$ denotes the edge from parent vertex $j$ to child vertex $i$ and $\mathcal{W}=(w_{ij})_{e_{ij}\in \mathcal{E}}$ with $w_{ij} \geq 0$ is the set of $m$ edges weights which are chosen by the community members in a heuristic manner. In our experiment, we used the weights according to Table \ref{table:graph_weights} where $w_{ij}$ is taken from \ref{table:edges_weights} if $v_i \neq s_0$ else \ref{table:nodes_weights}. The weights have a strong influence on the PageRank computation and the final reward, thus they must ensure that contributions that require a lot of labor get rewarded more than simpler ones and that they get rewarded when they are validated or reviewed by the community. Indeed, if a contribution $j$ has multiple children, its score should be propagated forward to the most important child.

\begin{figure}[h!]
	\centering
	\includegraphics[width=0.6\textwidth, keepaspectratio]{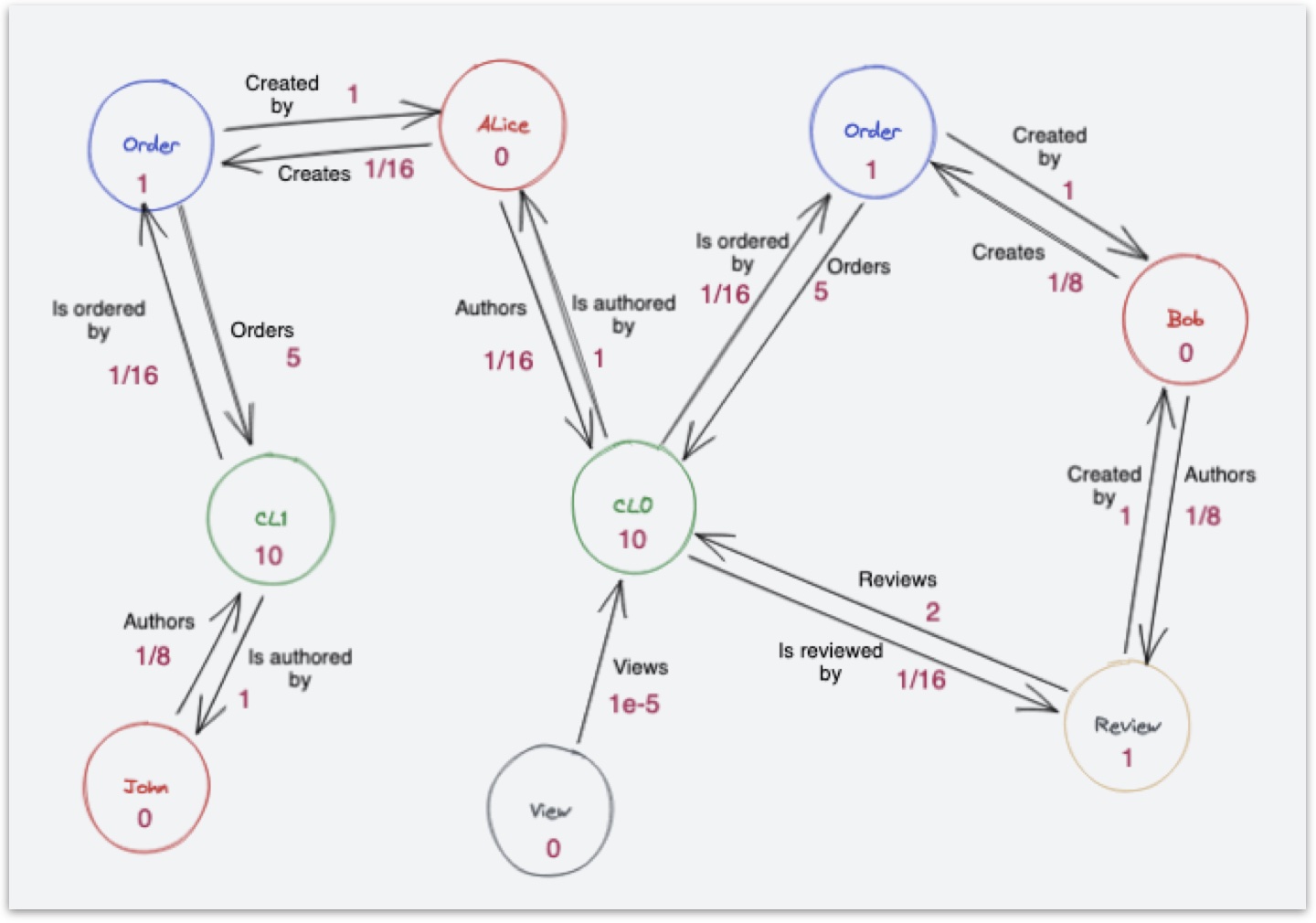}
	\caption{Contribution graph}
	\label{fig:contrigraph}
\end{figure}

Figure \ref{fig:contrigraph} shows such a graph, omitting $s_0$ for clarity, with 3 community members, John, Alice, and Bob, who are connected via the edges between their contributions and interactions. We can see that the rank of courselet $CL0$ will mostly be propagated to its author Alice ($1/(1+1/16+1/16)=0.89$).

A simple definition of the PageRank score of vertex $i$ is given in the original paper:
\begin{equation}
    pr(v_i)= \sum_{j=1}^{n} A_{i j} \frac{pr(v_j)}{\sum_{k=1}^{n} A_{k j}}
\end{equation}
where is $A$ the adjacency matrix of the graph $G$. To avoid rank sink, \citeauthor{page1999} developed the random surfer model, where from any node, the user either continues to a children node with probability $1-\alpha$ or the user jumps to a random node chosen based on the distribution in $s$ with probability $\alpha$. PageRank is then defined as the stationary distribution, $pr$, of a random walk, $pr_t$, on the graph $G$ by the equation:
\begin{equation}
pr_{t+1} = \alpha s + (1-\alpha)pr_t P
\label{eq:pagerank}
\end{equation}
where $pr_t$ is the row vector of vertex probabilities at time $t$, $s$ is the $n$-vector $(1/n, \ldots, 1/n)$, $pr$ is the row vector of PageRank score $pr = \left(pr(v_i)\right)_{1\leq i\leq n}$, $P$ is the transition probability matrix defined by $\forall 1\leq i,j \leq n$ $P(i,j) = \frac{w_{ij}}{d_i}$ where $d_i = \sum_j w_{ij}$ is the degree of $v_i$. The random surfer model ensures that the random walk on any graph $G$ is strongly connected which guarantees the existence of the PageRank scores (from the fundamental theorem of Markov chains). PageRank can be estimated \textit{locally} on large graphs \citep{NIPS2013_99bcfcd7} or globally using the power iteration method if the graph is small enough. Thanks to PageRank, we can easily evaluate each node score by using the score of all parent nodes, that way, for each researcher, having highly valuable parents improves its score more than having less important ones. This evaluation motivates Qr's members to produce research of high quality.

Since contributions such as review or order are not at the source of Qr’s value, the Personalized PageRank (PPR) is used by defining $s$ as $s = (s_0, 0, \ldots, 0) = (1, 0, \ldots, 0)$ where $s_0$ is a \textit{seed} node that corresponds to Qr itself or \href{https://quantinar.com/}{quantinar.com} homepage from the point of view of the random surfer model. 
In that case, it can easily be observed that for any node $v_i$, its PPR, $pr(v_i)$, is equal to the success probability that a random walk starting at $s_0$ and independently terminating at each time step with probability $\alpha$, hits $v_i$ just before termination 
\citep{doi:10.1080/15427951.2013.802752}. The PPR gives a contextualized ranking of Qr's contributions with respect to Qr's homepage.

\begin{figure}[h!]
	\centering
	\includegraphics[height=0.6\textwidth, keepaspectratio]{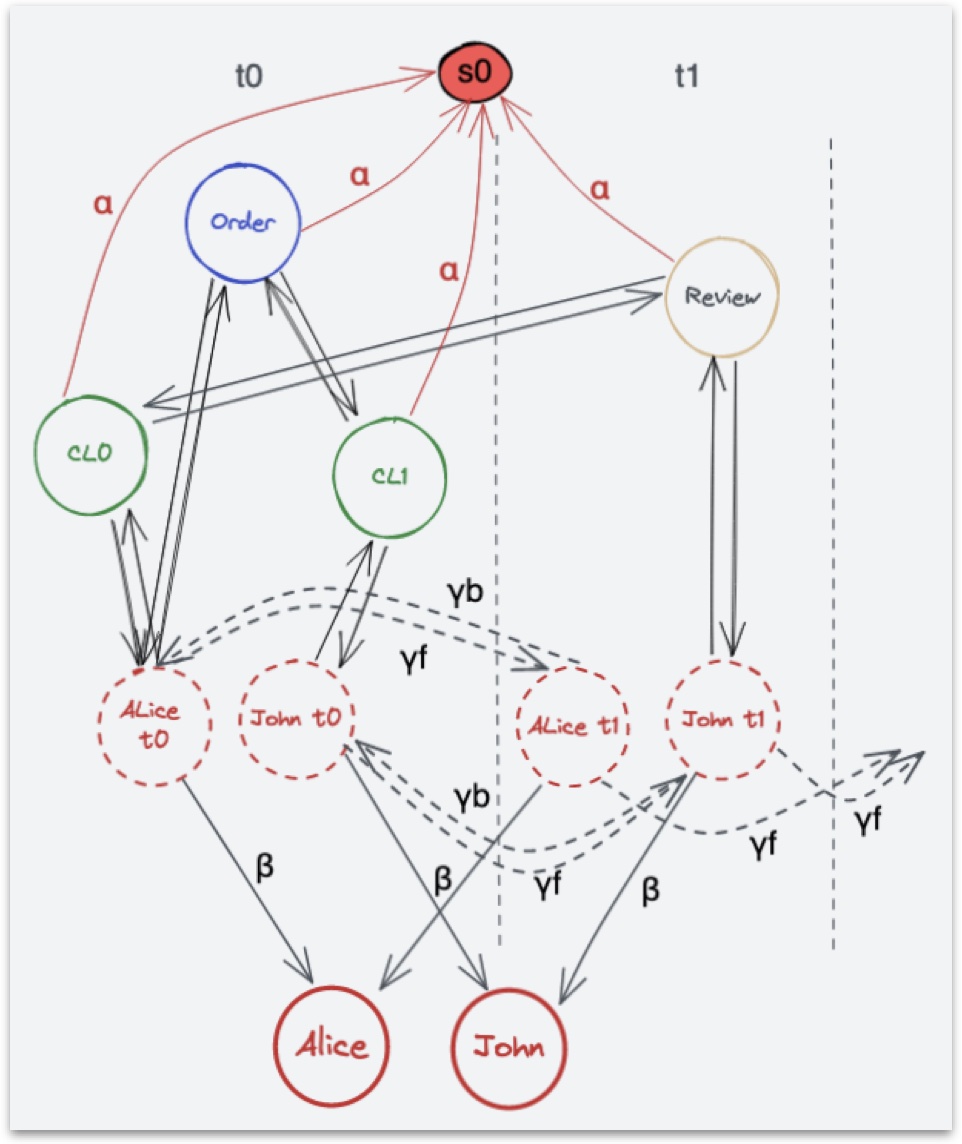}
	\caption{Contribution graph over periods}
	\label{fig:contrigraph_temp}
\end{figure}

Alas, the contribution graph defined above does allow us to study the PPR scores over time, in particular, the inertia of the score is not easily available. To overcome that issue, we propose to use the latest CredRank algorithm from SourceCred to evaluate the scores on a discretized historical graph as it is represented in Figure \ref{fig:contrigraph_temp}.

First, each vertex $i$ is indexed by a timestamp $t$ corresponding to its appearance in Qr and denoted as $v_{it}$. The history is then discretized in $T$ fixed length periods (for example a week). Let us denote $\mathcal{C}$ the set of vertices including all contributors and $\mathcal{A}$ the set of vertices including all contributions. We define $\mathcal{A}_k = \{v_{it},\ 1\leq i \leq n,\ 1\leq t \leq k,\ v_{it} \in \mathcal{A}\}$ as the set of vertices including all contributions until the date $k$. We can easily retrieve the set of new contributions during the period $k \geq 1$ with $\mathcal{A}_k^{new} = \mathcal{A}_k \setminus \mathcal{A}_{k-1}$ and $\mathcal{A}_0^{new} = \mathcal{A}_0$. At each period $k$, we create the epoch contributor nodes that authored the new contributions in $\mathcal{A}_k^{new}$, that is $\mathcal{C}_k^{new} = \{v_{ik},\ 1\leq i \leq n$ if $e_{ij} \in \mathcal{E}, v_i \in \mathcal{C},\ v_j \in \mathcal{A}_k^{new}\}$. Thus the set of epoch contributor nodes until period $k$ is defined as $\mathcal{C}_k = \bigcup_{1\leq t \leq k} \mathcal{C}_k^{new}$. We add directed edges from the contributions in $\mathcal{A}_k^{new}$ to their respective author(s) in $\mathcal{C}_k^{new}$. On top, webbing edges between previous and new epoch contributors are added with associated transition probability $\gamma_f$ and $\gamma_b$ connecting $\forall v_i \in \mathcal{C}_{k-1}$, $v_{ik-1}$ and $v_{ik}$ frontward and backward, respectively. The webbing edges ensure that the reputation gained at previous epochs has inertia over time. Finally, each epoch contributor node from $\mathcal{C}_k$ is connected to its associated contributor node in $\mathcal{C}$ with transition probability $\beta$. Thanks to this graph, each contribution is connected to its associated contributor only through the epoch contributor node and not the global contributor node. That way, for any contributor $i$ and period $k \geq 1$, the PageRank score at a previous period $k^{'} < k$ is easily accessible at the epoch contributor $v_{ik}'$.

For any $k,\ 1 \leq k \leq T$, we define the epoch contribution graph as $G_k = (\mathcal{V}_k, \mathcal{E}_k, \mathcal{W}_k)$ where $\mathcal{V}_k = \{s_0\} \cup \mathcal{A}_k \cup \mathcal{C}_k \cup \mathcal{C}$, $\mathcal{E}_k = \{e_{ij}\}_{(v_i, v_j) \in \mathcal{V}_k^2}$ and $\mathcal{W}_k=(w_{ij})_{e_{ij}\in \mathcal{E}_k}$. The graph update is then given by: $$G_{k}=\left(\mathcal{V}_{k-1}\cup \mathcal{V}_{k}^{new}, \mathcal{E}_{k-1} \cup \mathcal{E}_{k}^{new}, \mathcal{W}_{k-1} \cup \mathcal{W}_k^{new} \right)$$

Before evaluating PageRank on such a graph, we need to compute the transition probability $P$ to transform the contribution graph into a Markov chain. Given our definition, we have a mix of edges defined with weights and transition probability ($\alpha$, $\beta$, $\gamma_b$, or $\gamma_f$). For any node $v_i$, the transition probability associated with an outbound edge $e_{ij}$ is given if $v_j$ is an epoch contributor node or a contributor node or the seed node ($\gamma_b$ or $\gamma_f$, $\beta$, and $\alpha$, respectively). For other outbound edges, it is trivial to transform the associated weight to a transition probability that keeps relative outbound weights' importance by setting it to the percentage of the remaining transition probability with the formula: $P(ij) = \frac{w_{ij}}{\sum_j w_{ij}} (1 - c_i) $ where $c_i$ is the sum of the outbound predefined transition probability for node $v_i$, for example, $c_i = (\gamma_b + \gamma_f + \beta)$ if $v_i$ is an epoch node contributor or $c_i = \alpha$ if $v_i$ is a contribution. Having specific transition probabilities predefined allows us to keep a certain control on the reputation score dynamic. Indeed, from any contribution regardless of its outbound edges, the transition probability to the seed node is always equal to $\alpha$; similarly, from any epoch contributor node, regardless of its outbound edges in particular contributions, the transition probability to its global contributor node is always equal to $\beta$. The personalized PageRank scores, $pr(v_i)$ are obtained by solving \eqref{eq:pagerank} on $G_k$.

To compute the global score for any contributor $v_i \in \mathcal{C}$ at a given period $k$, a weighted sum that aggregates the previous epoch score with a discount factor, $r$, is used:
\begin{equation}\label{eq:score_discount}
    S_{ik}^{*} = \sum_{t=1}^k r^{k-t}pr(v_{it})
\end{equation}

Now, let us verify that $r$ acts as a control parameter over the inertia of the score over the evaluation periods. We can easily show that if there is no change in the contribution graph, $\forall k$, $S_{i,k+1}^{*} - S_{ik}^{*} = -(1-r)S_{ik}^{*} + \mathcal{O}(\gamma_f^k)$. Thus as $r$ decreases the inertia of the score over the previous period decreases. The inertia is maximized for $r=1$ and the score is roughly constant.

Since $pr(v_i)$ is a probability, $S_i^*$ defined above can be very small as $n$ increases. That is why SourceCred propose to normalize the score and scale it by the total amount of $\operatorname{QNAR}$ token minted at a given period. Any contribution node $j$ that has a strictly positive weight $w_{s_0j} > 0$ associated with the inbound edge from the seed node $s_0$ will mint $w_{s_0j}$ $\operatorname{QNAR}$ tokens. Let us denote $M_k=\sum_{j=1}^n w_{s_0j}$, the following definition of the reputation score, the QNAR score, for any contributor $i$ is then used:
\begin{equation}
    S_{i} = \frac{S_{i}^{*}}{\sum_{j \in \mathcal{C}} S_{j}^{*}}M_k
\label{eq:rep_score}
\end{equation}

With definition \eqref{eq:rep_score}, the reputation score will increase as the relative contributor page rank score and the total number of contributions that are valuable increase. Finally, this reputation system favors competition in the long term because to maximize one's reputation score, one needs to maximize the long-term value of one's contributions.

\subsubsection{A reputation based reward}\label{sec:reward}
SourceCred proposes three types of grain-generation strategies based on the QNAR scores: \textit{immediate, balanced} and \textit{recent}. The \textit{immediate} strategy mints one QLET, for each QNAR gained in the past week. The \textit{balanced} strategy mints QLET based on the weights that are given to each action inside the DAO at any moment. For example, if creating a courselet changed its weight from $w$ to $w'$, the QLET target will be recalculated for each author and they will be paid accordingly from the future payoff to match the new total QNAR score incorporating the weight change. The last strategy of generating QLET is \textit{recent}. This strategy refers to a discounted reward based on previous period QNAR scores.

Since the concept of a decentralized P2P platform is new, the Quantinar DAO will use the \textit{balanced} strategy. In the long run, this strategy is more equitable for all the contributors and allows for a continuous fine-tuning of weights used for QNAR/QLET generation by the whole community.

\subsection{QNAR dynamics}
Quantinar has already implemented certain features and collected some statistics. To illustrate the dynamics of the CredRank algorithm for the platform, we made an experiment using the production SQL dataset from the Quantinar platform and Google Analytics statistics of \href{https://quantinar.com/}{quantinar.com} from the inception date on 2021-09-13 to 2022-10-25. This dataset includes the following contributions:
\begin{itemize}
    \item CL publication as an author
    \item CL order (ordering a courselet as a user, in most cases the CLs are free)
    \item CL review (grading or commenting on a courselet as a user)
    \item CL page view
    \item Course publication as an author
\end{itemize}
The graph is created with the contributions \{"course", "courselet", "order", "review", "view"\} and user nodes. The edge weights and seed-to-contribution minting weights are defined respectively in Table \ref{table:edges_weights} and  \ref{table:nodes_weights}.

\begin{table}[!htb]
    \begin{subtable}{.5\columnwidth}
      \centering
        \begin{tabular*}{.5\columnwidth}{@{\extracolsep\fill}ll@{\extracolsep\fill}}
            \toprule
            Edge & Weight \\
            \midrule
            (view, courselet) & 1e-5 \\
            (course, user) & 1 \\
            (user, course) & 1/8 \\
            (courselet, user) & 1 \\
            (user, courselet) & 1/8 \\
            (course, courselet) & 1 \\
            (courselet, course) & 1/8 \\
            (order, course) & 5 \\
            (course, order) & 1/16 \\
            (user, order) & 1/16 \\
            (order, user) & 1 \\
            (user, review) & 1/8 \\
            (review, user) & 1 \\
            (review, course) & 2 \\
            (course, review) & 1/16 \\
            \bottomrule
        \end{tabular*}
        \caption{Edges weights \label{table:edges_weights}}
    \end{subtable}%
    \begin{subtable}{.5\columnwidth}
      \centering
        \begin{tabular*}{.5\columnwidth}{@{\extracolsep\fill}ll@{\extracolsep\fill}}
            \toprule
            Edge & Weight \\
            \midrule
            courselet & 8 \\
            review & 1 \\
            order & 1 \\
            course & 0 \\
            view & 0 \\
            user & 0 \\
            \bottomrule
        \end{tabular*}
        \caption{Seed-to-contribution weights \label{table:nodes_weights}}
    \end{subtable}
    \caption{Graph weights}
    \label{table:graph_weights}
\end{table}. It is worth noticing that publishing a course does not mint any QNAR. Indeed, as courses contain courselet(s), only publishing a courselet should mint additional QNAR. On top, as Quantinar offers the users to create courses from courselet(s) that they did not author, a possible malicious behavior could appear: creating courses that contain courselet(s) with high reputation to steal it from the original courselet and artificially increase one's QNAR score. A simple mechanism to avoid this issue would be to add a new node for any linked courselet that does not have any inbound edge from the original one, but one outbound edge from the linked courselet to its original. That way, the instant reward of linking a courselet can only be positive for the original author.

Finally, we define the following transition probabilities using default values from SourceCred: $\alpha = 0.1$, $\beta = 0.4$, and $\gamma_f = \gamma_b = 0.1$. This choice is governed by the aim to guarantee that the score of each contribution is influenced by earlier dependencies and that the score of each contributor has some inertia over time. The effect of those parameters on the reputation score has been studied by the SourceCred team (\href{https://hackmd.io/@mzargham/SkY7VvQnV?type=view\#Contextualizing-Cred}{Contextualizing Cred}, \href{https://github.com/sourcecred/research}{github.com/sourcecred/research}) and it is future research in the context of Quantinar. However, their impact on the QNAR score can be described analytically. 

Indeed, for any node $v_i$ in the graph, the contribution from a connected parent $v_j$ with score $pr(v_j)$ and $n-1$ nodes between $v_i$ and $v_j$, is bounded by $(1-\alpha)^n pr(v_j)$. $1-\alpha$ corresponds to the rate at which the contribution value propagates itself to neighbors in the graph. Thus, decreasing $\alpha$ shifts the credit from the most immediate to further children's contributions. A small $\alpha$ is required if one attributes more value to the most foundational work.

$\gamma_f$ and $\gamma_b$ influences mostly the contributors reputation. Indeed, we can easily show that for small $\gamma_f$ and $\gamma_b$, $\exists M>0$ and $\forall t > M$, the score of the epoch contributor node $v_{it}$ is given by $pr(v_{it})=\frac{1}{1-\gamma_f\gamma_b}\left( C_{it} + \gamma_f C_{it-1} + \gamma_f^2 C_{it-2} \right) + \mathcal{O}(\gamma_f^3)$, where $\mathcal{A}_t^{new}$ is the set of new contributions at period $t$ and $C_{it} = \sum_{v_j \in \mathcal{A}_t^{new}} \pi_{ji} pr(v_j)$, the share of the epoch contributor score that comes from its new contributions only. We provide a short demonstration:

At epoch $t$, we have $pr(v_{it}) = \gamma_fpr(v_{i,t-1}) + \sum_{v_j \in \mathcal{A}_t^{new}} \pi_{ji} pr(v_j)$, where $v_{it}$ is an epoch contributor node and $\mathcal{A}_t^{new}$ is the set of new contribution at $t$. Developing the recurrence to previous epoch nodes, we have $\forall t > k,\ \forall k \geq 1,\ pr(v_i)^{t-k} = \gamma_fpr(v_{i,t-k-1}) + \gamma_b pr(v_{i,t-k+1}) + C_{it-k}$, thus $\forall t \geq 3,\ pr(v_{it}) = \frac{1}{1-\gamma_f\gamma_b} \Bigl \{\gamma_f^t pr(v_{i0}) + \gamma_b \sum_{k=0}^{t-3}\gamma_f^{k+2}pr(v_{i,t-k-1}) + \sum_{k=0}^{t-1}\gamma_f^k C_{it-k} \Bigr \}$. The result is derived using $\gamma_f < 1$.

Thus, $\gamma_f$ controls the inertia of the previous epoch scores within an evaluation period. Increasing $\gamma_f$ shifts the value from the most recent to older contributions.

\begin{figure}[h!]
	\centering
	\includegraphics[width=0.6\textwidth, keepaspectratio]{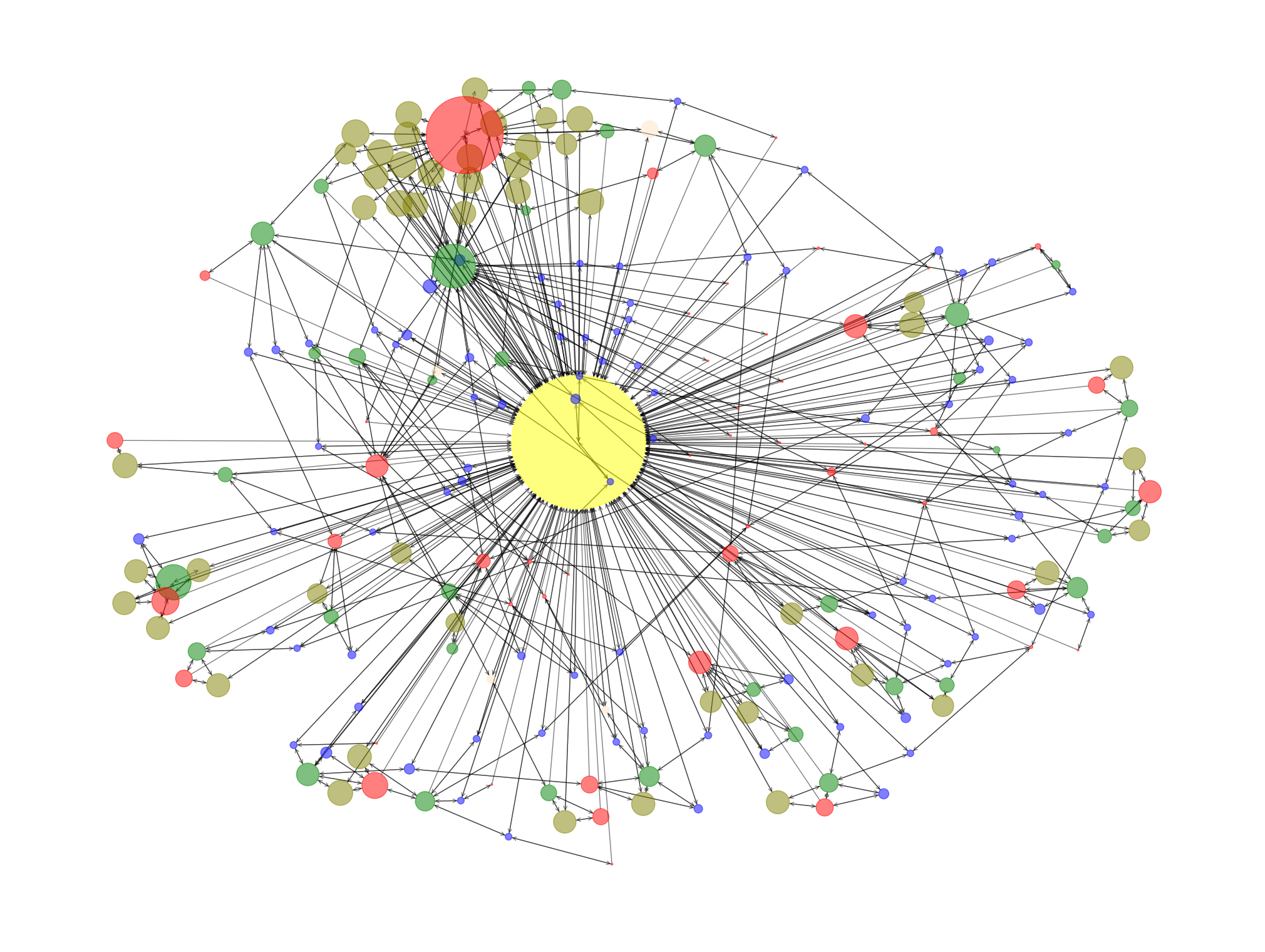}
    \captionsetup{width=0.6\textwidth}
    \caption{PageRank Contribution graph at 2021-10-19: \textcolor{seed-node}{seed}, \textcolor{author-node}{contributor}, \textcolor{course-node}{course}, \textcolor{courselet-node}{courselet}, \textcolor{review-node}{review}, \textcolor{order-node}{order}. The node size is proportional to the associated PageRank score \href{https://github.com/QuantLet/quantinar-rep}{\quantlet Quantinar-rep}}
	\label{fig:pr-contri-graph}
\end{figure}

\begin{figure}[h!]
	\centering
	\includegraphics[width=0.6\textwidth]{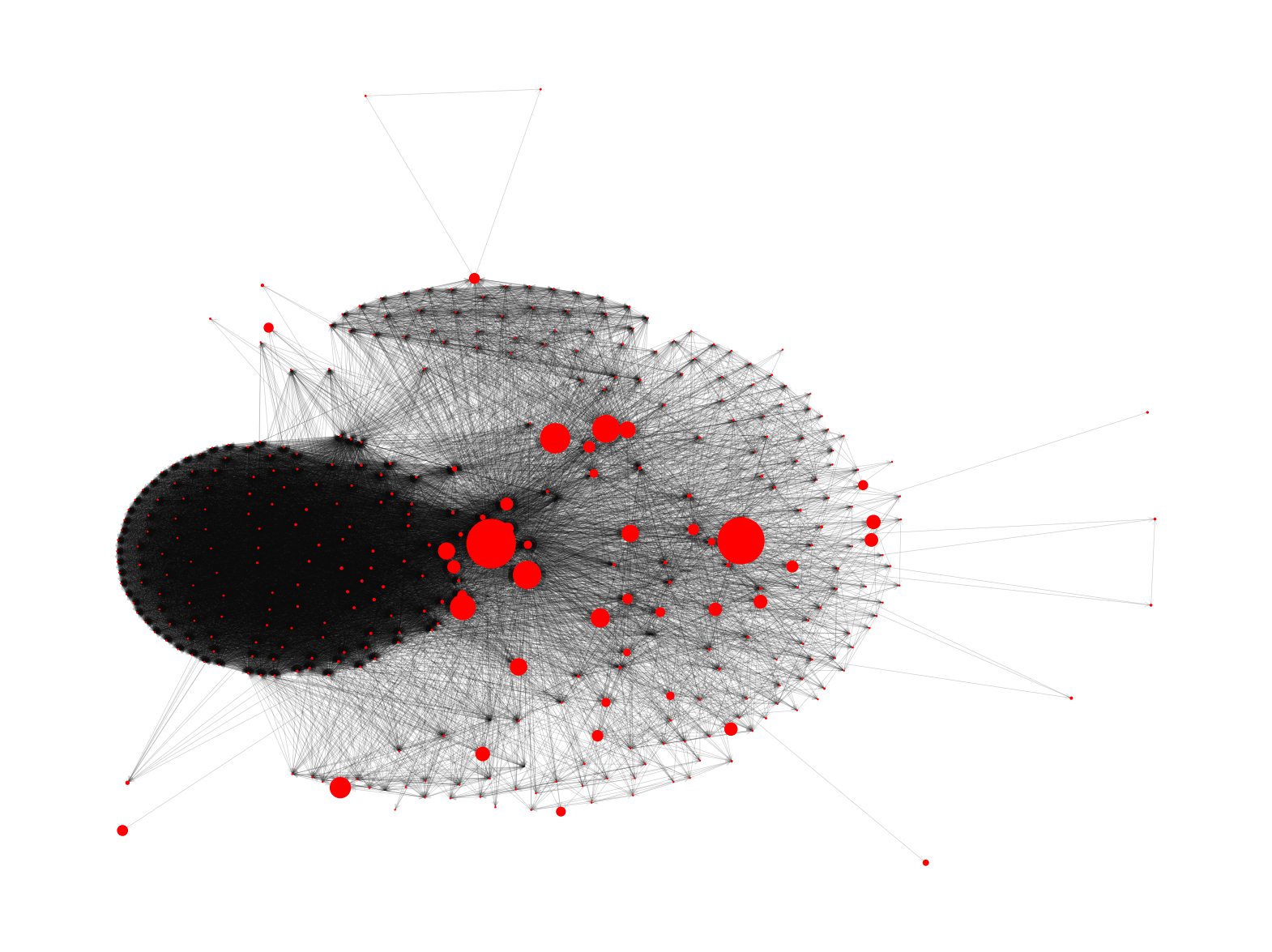}
    \caption{Contributor subgraph at 2022-10-18. The node size is proportional to the associated QNAR score. An edge is created between two users if they are linked through at least one contribution \href{https://github.com/QuantLet/quantinar-rep}{\quantlet Quantinar-rep}}
	\label{fig:pr-author-graph}
\end{figure}

Before looking at the QNAR score, an example of the contribution graph at 2021-10-12 is presented in Figure \ref{fig:pr-contri-graph}.  We can observe one contributor that, at that time, participates the most in the Q2 ecosystem as the user publishes multiple CLs into various courses that are ordered multiple times by different users. By doing so, the user increases its Pagerank score, which is visualized with a larger node size in the figure. Figure \ref{fig:pr-author-graph} shows the latest contributors' sub-graph with their associated QNAR score on 2022-10-18, as the full contribution graph is too large at that date. Two contributors act as central members and have the largest QNAR score. We can also observe a community within Quantinar at the left of the graph which seems tightly influenced by the left central node on the graph. 

\begin{figure}[h!]
	\centering
	\begin{subfigure}[b]{0.45\textwidth}
		\centering
		\includegraphics[width=\textwidth]{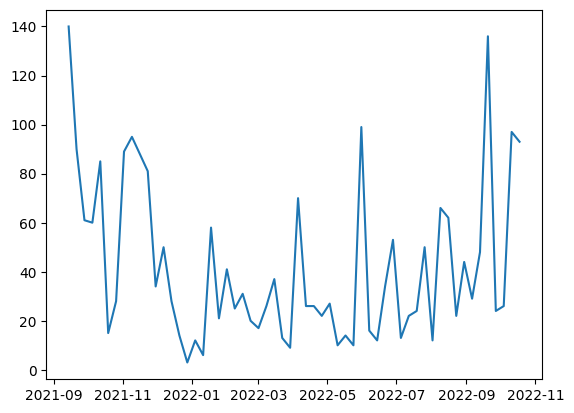}
		\caption{Minted QNAR per weekly period}
		\label{fig:minted_qnar}
	\end{subfigure}
	\hfill
	\begin{subfigure}[b]{0.45\textwidth}
		\centering
		\includegraphics[width=\textwidth]{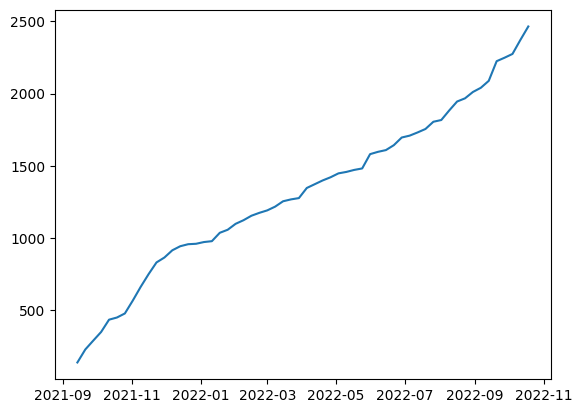}
		\caption{Cumulative minted QNAR}
		\label{fig:cum_minted_qnar}
	\end{subfigure}
    \captionsetup{width=0.9\textwidth}
    \caption{QNAR supply dynamics \quantlet \href{https://github.com/QuantLet/quantinar-rep}{Quantinar-rep}}
	\label{fig:qnar_dynamics}
\end{figure}

Using the temporal graph presented in the previous paragraph, we evaluate the QNAR score using a slow decay rate ($r=0.95$). \ref{fig:minted_qnar} and \ref{fig:cum_minted_qnar} show the weekly minted QNAR and cumulative minted QNAR, respectively. In Figure \ref{fig:minted_qnar}, we observe that Qr alternates between a period of high productivity and a period where new contributions are less frequent. This behavior justifies the use of the temporal graph since periods do not have the same importance. Finally, Qr's value, measured in terms of QNAR, presents an up-trend since the beginning of 2022, which has strengthened in the second half of 2022, showing relative traction for Quantinar. 

\begin{figure}[h!]
    \centering
    \includegraphics[width=0.6\textwidth]{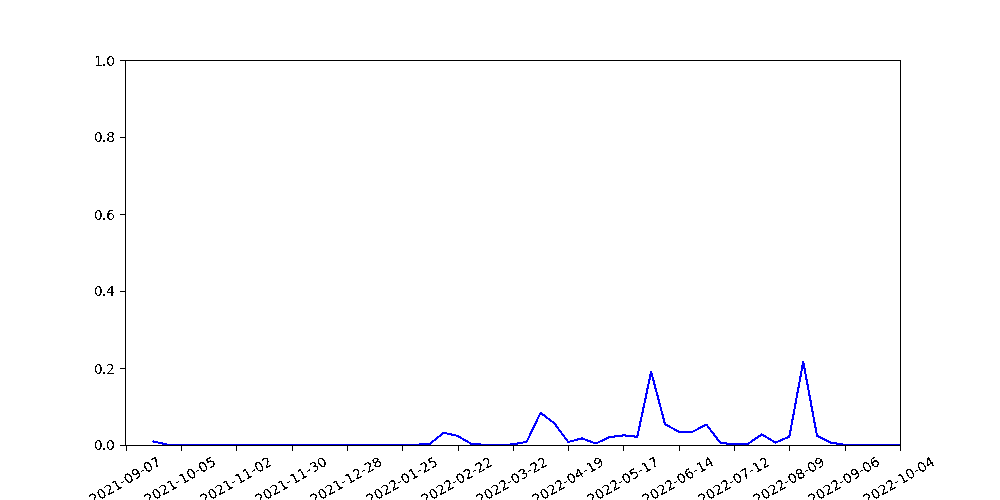}
    \caption{Example of epoch relative importance in QNAR score (in \% of the total score).\\ The relative epoch importance is presented for the top user at the latest date \quantlet \href{https://github.com/QuantLet/quantinar-rep}{Quantinar-rep}}
    \label{fig:epoch_importance}
\end{figure}

On top, we can observe in Figure \ref{fig:epoch_importance} that the score behaves as expected for the top user: the score on 2022-10-18 is mostly influenced by the contributions from previous periods. As past contributions become more popular, the share of the past epoch nodes' score within the total score will increase, which motivates community members to appreciate the long-term value of their contributions. Indeed, the weeks 2022-03-29, 2022-05-31, and 2022-08-23 are the most important for the user on 2022-10-18 and contribute roughly 10, 20, and 22\% in the final score, respectively.

\begin{figure}[h!]
	\centering
	\begin{subfigure}[b]{0.45\textwidth}
		\centering
		\includegraphics[width=\textwidth]{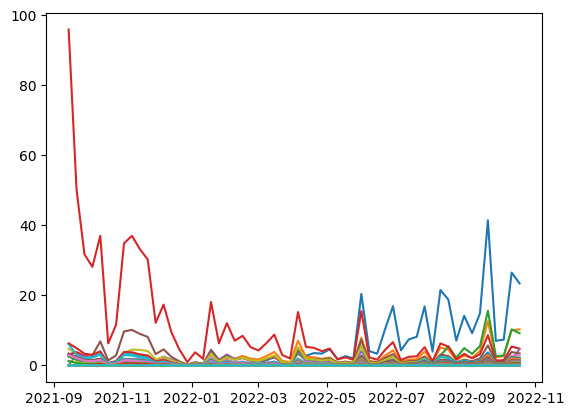}
		\caption{QNAR user score}
		\label{fig:user_qnar}
	\end{subfigure}
	\hfill
	\begin{subfigure}[b]{0.45\textwidth}
		\centering
		\includegraphics[width=\textwidth]{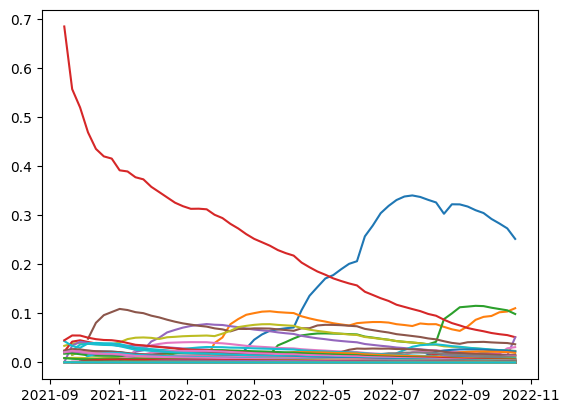}
		\caption{Relative QNAR user score}
	    \label{fig:user_qnar_rel}
	\end{subfigure}

    \captionsetup{width=0.9\textwidth}
    \caption{QNAR score dynamics using a weekly period and discount rate $r=0.95$ \quantlet \href{https://github.com/QuantLet/quantinar-rep}{Quantinar-rep}}
\end{figure}

Figure \ref{fig:user_qnar} shows the final weekly QNAR scores per user and Figure \ref{fig:user_qnar_rel} their relative score. We observe that the score behaves as expected. The QNAR has inertia over time which guarantees that the contributions from the previous period still have values today. Nevertheless, contributors need to consistently add contributions to keep up with new active contributors, which can be observed when comparing the red and blue curves in Figure \ref{fig:user_qnar_rel}. Indeed, as the evaluation period is more recent, the score of the early contributor in red decreases at an inversely exponential rate from September 2021, as the user probably reduced its contribution to Quantinar. On the other hand, the user corresponding to the blue curve starts to be very active in March/April 2022 and his score increases drastically until he reached the highest score in mid-July 2022.

At the latest evaluation date, the top 4 users in terms of CL creation also belong to the top 4 users in terms of QNAR reward which demonstrates that Qr rewards researchers first. However, the user corresponding to the red curve is the top contributor in terms of CL creations (see Figure \ref{fig:courselet_users}), but only the 4th in terms of QNAR reward, as its important contributions are too old to be the most valuable in recent periods thanks to the discount factor $r$ in \eqref{eq:score_discount}.

Finally, Figure \ref{fig:user_qnar} shows that the user score is quite volatile, as it is a function of the periodic minted QNAR presented in Figure \ref{fig:minted_qnar}. We could propose to smooth it to have a more stable score across evaluation periods.

\subsection{An auction based open peer-review (OPR)}\label{sec:opr}

An open peer-review process using IPFS and decentralization has already been conceptualized by \citep{tenorioetal:2019}. Such a process ensures the seven traits of the open peer-review process identified by \citet{Ross-Hellauer:2017}:
\begin{itemize}
    \item \textbf{Open identities}: Authors and reviewers are aware of each other’s identities.
    \item \textbf{Open reports}: Review reports are published alongside the CL and versioned using IPFS.
    \item \textbf{Open participation}: The wider community can contribute to the review process. This is ensured by an open call for review from the CL's author.
    \item \textbf{Open interaction}: Direct reciprocal discussion between author(s) and reviewers, and/or between reviewers, is allowed and encouraged. Since the authors' and reviewers' identities are revealed to each other, they are encouraged to chat in a specific Discord channel which is open for reading the community.
    \item \textbf{Open pre-review manuscripts}: CLs are made immediately available on Quantinar in advance of any formal peer review procedures.
    \item \textbf{Open final-version commenting}: Review or commenting on final publications is allowed directly on the CL page on Quantinar.
    \item \textbf{Open platforms}: Quantinar is open by design.
\end{itemize}

In Quantinar, on top of addressing the above issues, we propose a new peer-review process that motivates community members to review publications in a time-effective and fair manner. 

Let there be a smart contract, a computer program intended to automatically govern and execute the rules of the following game and manage the associated cash flow. The game is defined with stake owner ${S_i}$ with stakes ${s_i}$, paper proposer $P$, paper acceptance status $S$ (Accepted, Denied), auction type $T$ maturity $M$, payoff function $f$, voting function $v$, token $\operatorname{QNAR}$, inflation rate $I$. The goal of the game is to find a majority-driven and fair peer review for a proposed paper and to encourage publication. While the latter can be achieved through token rewards, due process is designed for the former. A majority-driven and fair review is to be defined as one that is simultaneously qualified and aligned with the majority of the network. Since the majority of decision-makers are not necessarily objectively right, an incentive structure must be created that assigns a larger weight to those decision-makers, who have a history of being right or publish knowledge that is deemed right (this assessment can be made with the help of the reputation score, see Section \ref{sec:reputation}). 

The inflation rate, time to maturity, auction type, and payoff function are defined as global variables, whose status can be set by the DAO. The payoff function is in its simplest form a function that sums all bids and assigns a positive (negative) bid-weighted share to each winner (loser). Additionally, each participant receives token units according to the inflation rate after each iteration. A multitude of auction types is possible. Auctions are conducted in terms of a First-price sealed bid, implemented in a smart contract \citep{blindAuctionContract}. Other classes of auctions might be employed in the future.

Let there be a proposal for a new paper by proposer P. To counteract spam and incentivize reviewer participation, the proposer must submit a small initial bid. Yet the proposer is excluded from reviewing his/her paper, thus only profiting if the paper is accepted and adds value to the network. The acceptance of the proposal is treated as an auction that inherits above defined global variables. Stake Owners $S_1, S_2, S_3$, of Token QNAR are allowed to participate in the auction. Let's assume that each Stake Owner enters the auction as a player. 
Having reviewed the paper each on their own, the players must decide on the Acceptance Status S in a blind auction. They submit a bid, which is a subset of their stake, encrypted with their public key and their voting decision (Accepted or Denied) as input to a smart contract. When the auction closes at maturity, all players must publicly disclose their bid. Since their public keys are known, verification of their bid is trivial. In case of a false claim, the player automatically loses their (locked up) bid. Due to the strict negative payoff, there cannot be any incentive to make a false claim.

$$
\begin{aligned}
r = \frac{\sum_{j \in \mathcal{L}}{s_j}}{\sum_{i \in \mathcal{W}}{s_i}} \\
P(S_i) = \frac{s_i}{\sum_{j \in \mathcal{W}}{s_j}} * r \\
\end{aligned}
$$

where $P(S_i)$ is defined as the profit of Staker $S_i$. The index $\mathcal{L}$ and $\mathcal{W}$ correspond to stakes on a game's losing or winning side. The ratio of the sum of stakes on either side of the bet balances diminishes the bet's payoff if a structural imbalance exists. 

As an example, use a voting function $v$ for the acceptance status, e.g. meaning that "Accepted" is assigned 1 and "Denied" is assigned -1. Then, at the close of the auction, the voting decision is defined according to the sign of the bid-weighted average of votes. 
Let player $S_1, S_2$ vote "Accepted" with a stake of 1 and 2 $\operatorname{QNAR}$s, whereas player $S_3$ votes "Denied" with 2 $\operatorname{QNAR}$s. 
Then the sum of bids is 5 $\operatorname{QNAR}$s. The voting decision is $1 + 2 - 2 = 1$. The potential payoff for the winners is restricted by the ratio of stakes on each side of the game. This is an optional function that aims to dampen the effects of a strong imbalance of betting forces. It achieves this goal by decreasing the marginal payoff for each Staker to join the stronger side. The sum of bids is then distributed proportionally to the players $S_1$, who receives $(1/3) \times 5$, and $S_2$, who receives $(2/3) \times 5$. 

The rate of minted tokens determines the general degree of incentive that potential reviewers have; which is independent of their abilities. Since only people who participate in reviews receive additional tokens, non-participants bear the cost of inflation. Hence the rate of minted tokens should be decided by the network.
To prevent proposers to employ trivial strategies to profit from minting, an anti-spam strategy needs to be enforced. We suggest that each proposer is required to submit a minimum bid with each paper. The bid size should also be decided by the network. 

\subsubsection{Simulation}

A simulation study is conducted to assess the minimum amount of participants required for a stable system, their time of survival, and expected changes in wealth (stakes) conditional on the initial distributions.

An artificial paper acceptance probability is introduced and assumed to be 0.5. Inflation per iteration is 1 QNAR.
The aggregate behavior of stakers is simulated. A staker's acceptance or denial of a paper is a Bernoulli-distributed random variable with 0.5 probability. For each type of initial stake distribution, Pareto and uniform, stakers performances are evaluated for a set of 5, 10, 50, 100, and 1000 participants after the rounds 10, 50, 100, 1000, and 10000. One round is counted as the proposal of a paper, meaning the submission and subsequent acceptance or denial of a paper. 

\begin{figure}[h!]
	\centering
	\begin{subfigure}[b]{0.45\textwidth}
		\centering
		\includegraphics[width=\textwidth]{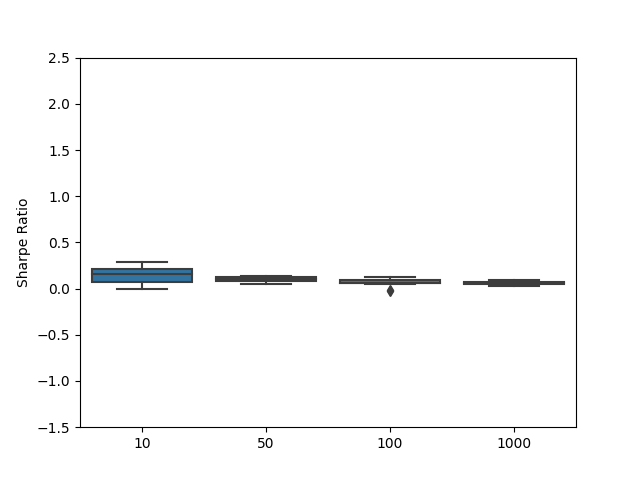}
		\caption{Initial Pareto Distribution}
		\label{fig:boxthree}
	\end{subfigure}
	\hfill
	\begin{subfigure}[b]{0.45\textwidth}
		\centering
		\includegraphics[width=\textwidth]{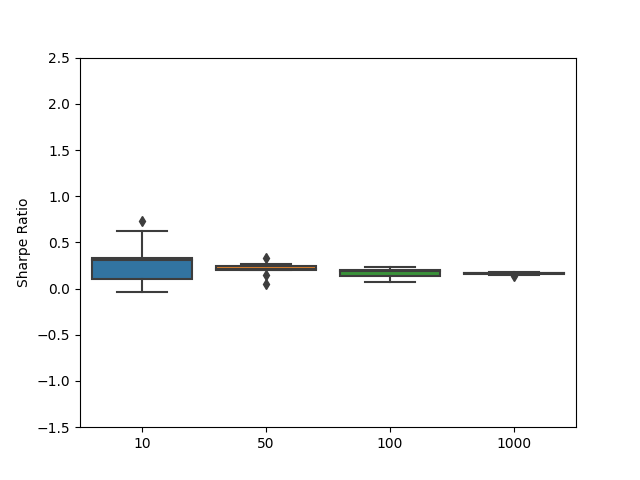}
		\caption{Initial Uniform Distribution}
		\label{fig:boxfour}
	\end{subfigure}
	\caption{Sharpe Ratio conditional on amount of Stakers \quantlet \href{https://github.com/QuantLet/Quantinar-Staking-Simulation}{Quantinar-Staking-Simulation}}
	\label{fig:boxplot_staker}
\end{figure}

We find the system to be stable for as few as 5 players. Figure \ref{fig:boxplot_staker} shows that Sharpe Ratios are generally high and positive when there are few stakers. Sharpe Ratios decrease with growing competition, yet stay positive even for a large number of stakers. This behavior is independent of the initial distribution, whether it is uniform- or Pareto-distributed. Consequently, reviewers are incentivized to offer their expertise especially when there is a lack of it because that is when their potential payoff is the highest. This incentive holds under real-world, uneven distributions such as of the Pareto type. Table \ref{table:table_performance} shows the expected return, standard deviation, and Sharpe Ratio per round, i.e. per paper proposal conditional on a different type of initial distribution. Under both types of distribution, stakers are strongly encouraged to participate due to the high expected return. While stakers generally earn less under an initial Pareto-distributed stake than under a uniform distribution, Sharpe Ratios and expected returns remain high.

\begin{figure}[h!]
	\centering
	\begin{subfigure}[b]{0.45\textwidth}
		\centering
		\includegraphics[width=\textwidth]{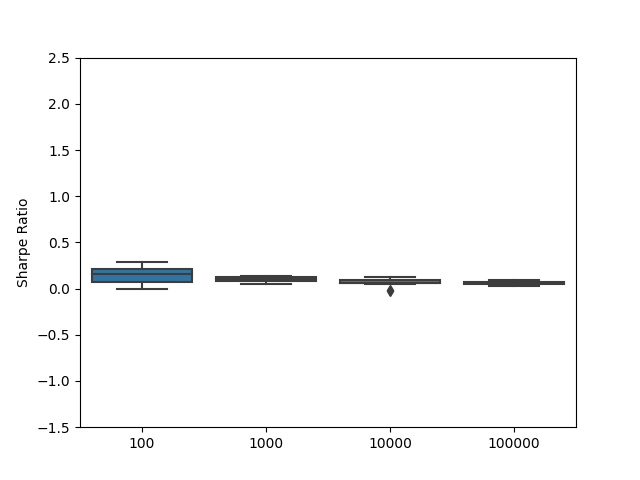}
		\caption{Initial Pareto Distribution}
		\label{fig:boxone}
	\end{subfigure}
	\hfill
	\begin{subfigure}[b]{0.45\textwidth}
		\centering
		\includegraphics[width=\textwidth]{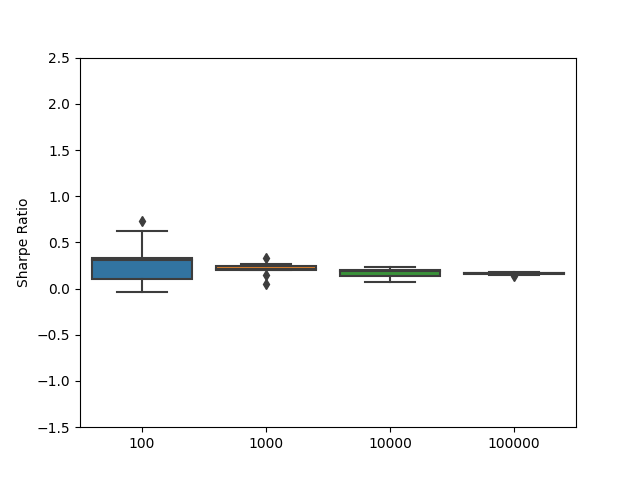}
		\caption{Initial Uniform Distribution}
		\label{fig:boxtwo}
	\end{subfigure}
	\caption{Sharpe Ratio conditional on amount of Paper Proposals \quantlet \href{https://github.com/QuantLet/Quantinar-Staking-Simulation}{Quantinar-Staking-Simulation}}
	\label{fig:boxplot_paper}
\end{figure}

\begin{table}[h!]
    \centering
    \begin{tabular*}{\columnwidth}{@{\extracolsep\fill}llll@{\extracolsep\fill}}
    \toprule
    Initial Distribution & Expected Return & Standard Deviation & Sharpe Ratio \\
    \midrule
    Uniform & 0.1009 & 0.5356 & 0.2057 \\ 
    Pareto &  0.0317 & 0.3969 & 0.0913 \\ 
    \bottomrule
    \end{tabular*}
    \caption{Performance Measures conditional on initial Wealth Distribution per proposed paper}
    \label{table:table_performance}
\end{table}

\section{Conclusion}

We propose a P2P platform, Quantinar, designed to accompany all players in the education and research system in their knowledge accumulation. Quantinar incentivizes academic publications based on an open and decentralized infrastructure that favors research findings' transparency and reproducibility. Quantinar is intended for students, practitioners, and researchers alike. We cater to the need for sound statistical and quantitative education in the modern world but do not restrict the educational themes on the platform. We aim to encourage a new scientific publication standard consisting of a triptych of CL, DL, and QL (knowledge, data, and code), allowing for vertical learning and ensuring the research findings' reproducibility. Quantinar members interact with other user-generated content within the C5 framework to create a consumer-prosumer spiral and encourage communication between researchers. Consumers and prosumers are motivated to participate via a token-based reward, which reflects their interactions within and contributions to the platform thanks to SourceCred's CredRank, a flexible data-driven reputation system inspired by the personalized PageRank algorithm. This reputation system ensures that the community decides which contributions to reward and that users maximizing the long-term value of their contributions obtains more rewards than shortsighted publications. Finally, we propose a new open peer-review process to assess the quality of submitted content. Independent reviewers are allowed to review and vote on paper proposals. The voters' stake partly supports the voting process, which facilitates honest reviews, counteracts spam, and incentivizes reviewers. This process reduces the presence of malpractices, such as a positive publication bias, to which traditional journals might be converging without radical changes.

\bibliography{reference}

\end{document}